\documentclass[twocolumn]{article}
\usepackage[margin=.8in]{geometry}
\usepackage{natbib}
\usepackage{amsmath}
\usepackage{amssymb}
\usepackage{graphicx}
\setkeys{Gin}{width=\linewidth}
\usepackage{mleftright}
\usepackage{tikz}
\usetikzlibrary{positioning}
\usepackage{fontawesome5}
\usepackage{comment}
\usepackage{float}
\usepackage{times}

\author{
Carlos Martin\\
cgmartin@cs.cmu.edu\\
Carnegie Mellon University
\and
Tuomas Sandholm\\
sandholm@cs.cmu.edu\\
Carnegie Mellon University\\
Strategy Robot, Inc.\\
Optimized Markets, Inc.\\
Strategic Machine, Inc.
}

\date{}

\newcommand{\diff}{\mathop{}\!\mathrm{d}}
\DeclareMathOperator{\expect}{\mathbb{E}}
\DeclareMathOperator{\argmax}{argmax}

\DeclareMathOperator*{\esssup}{ess\,sup}

\title{Solving Infinite-Player Games with Player-to-Strategy Networks}
\begin{document}
\maketitle
\begin{abstract}
We present a new approach to solving games with a countably or uncountably infinite number of players.
Such games are often used to model multiagent systems with a large number of agents.
The latter are frequently encountered in economics, financial markets, crowd dynamics, congestion analysis, epidemiology, and population ecology, among other fields.
Our two primary contributions are as follows.
First, we present a way to represent strategy profiles for an infinite number of players, which we name a \emph{Player-to-Strategy Network (P2SN)}.
Such a network maps players to strategies, and exploits the generalization capabilities of neural networks to learn across an infinite number of inputs (players) simultaneously.
Second, we present an algorithm, which we name \emph{Shared-Parameter Simultaneous Gradient (SPSG)}, for training such a network, with the goal of finding an approximate Nash equilibrium.
This algorithm generalizes simultaneous gradient ascent and its variants, which are classical equilibrium-seeking dynamics used for multiagent reinforcement learning.
We test our approach on infinite-player games and observe its convergence to approximate Nash equilibria.
Our method can handle games with infinitely many states, infinitely many players, infinitely many actions (and mixed strategies on them), and discontinuous utility functions.
\end{abstract}

\section{Introduction}

Many fields of research study real-world situations involving large numbers of agents.
These include economics, financial markets, crowd dynamics, congestion analysis, epidemiology, and population ecology.
The presence of many agents with their own individual interests---competing or otherwise---raises the challenge of how to model and analyze such systems.
It is often advantageous to model multiagent systems with a large number of agents as infinite-player games, because the limit of infinite players simplifies analysis.
This simplification is similar to the treatment of many-particle systems in thermodynamics and statistical mechanics.
One example of this in the context of game theory is \citet{Ganzfried10:Computing}, who presented a procedure for solving large imperfect-information games by solving an infinite approximation of the original game and mapping the equilibrium to a strategy profile in the original game.
Perhaps counter-intuitively, it is often the case that the infinite approximation can be solved much more easily than the finite game.

Infinite-player games present the challenges of (1) how to represent strategy profiles involving infinitely many players and (2) how to find strategy profiles that constitute an approximate Nash equilibrium, which is the standard solution concept in game theory.

We present a way to represent strategy profiles for an infinite number of players, which we coin a \emph{Player-to-Strategy Network (P2SN)}.
As its name suggests, such a network maps players to strategies.
It exploits the generalization capabilities of neural networks to learn across an infinite number of inputs (players) simultaneously.

To train such a network, we present an algorithm which we coin \emph{Shared-Parameter Simultaneous Gradient (SPSG)}.
It generalizes simultaneous gradient ascent and its variants, which are classical equilibrium-seeking dynamics used for multiagent reinforcement learning, to the case where a single set of parameters is used to represent many agents---in fact, infinitely many.

We test our approach on infinite-player games and observe its convergence to approximate Nash equilibria.
Our method can handle games with infinitely many states, infinitely many players, infinitely many actions (and mixed strategies on them), and discontinuous utility functions.

The rest of the paper is structured as follows.
In \S\ref{sec:related}, we describe related research.
In \S\ref{sec:formulation}, we introduce relevant notation and present a mathematical formulation of the problem we are solving.
In \S\ref{sec:optimization}, we describe the optimization techniques that our method leverages.
In \S\ref{sec:method}, we present our method itself.
In \S\ref{sec:experiments}, we present our experimental settings, results, and discussion.
In \S\ref{sec:conclusion}, we present our conclusions.

\section{Related Research}
\label{sec:related}

In this section, we briefly review some major results in the literature pertaining to infinite-player games, including theoretical properties as well as proposed methods for solving them.
One of the earliest papers on infinite-player games is by \citet{aumann1964markets}, who suggested that the most appropriate model of a market with many individually insignificant traders is one with a continuum of traders (like the continuum of points on a line).

\subsection{Equilibrium Existence}

\citet{schmeidler1973equilibrium} generalized Nash's theorem~\citep{Nash51:Non} on equilibrium existence to the case of a continuum of players endowed with a non-atomic measure.
\citet{khan1985equilibrium} generalized this result to non-atomic games in which each player's strategy set is a weakly compact, convex subset of a separable Banach space whose dual has the Radon--Nikodym property.
\citet{khan1986equilibrium} generalized this result to non-atomic games with strategy sets in a Banach space.

\citet{khan1987cournotcontinuum} proved the existence of a Cournot--Nash equilibrium in generalized qualitative games with a continuum of players, under certain conditions.
Similarly, \citet{khan1987cournotatomless} proved the existence of a Cournot--Nash equilibrium in games with an atomless measure space of players, each with unordered preferences and strategy sets in a separable Banach space.

\citet{kim1989equilibria} proved the existence of a Nash equilibrium for an abstract economy with a measure space of agents, infinite-dimensional strategy space, and agent preferences that need not be ordered.

\citet{rath1992direct} gave a simple proof of the existence of pure-strategy Nash equilibria in games with a continuum of players when a player's payoff depends only on its own action and the average action of others.
This was extended to the case where the action set of each player is a compact subset of \(\mathbb{R}^n\).
\citet{khan2002non} surveyed games with many players, reporting results on the existence of pure-strategy Nash equilibria in games with an atomless continuum of players, each with an action set that is not necessarily finite.

\citet{wiszniewska2014open} studied open- and closed-loop Nash equilibria in games with a continuum of players, and both private and global state variables, proving an equivalence theorem between these classes of equilibria.

\subsection{Equilibrium Uniqueness}

\citet{milchtaich1996generic} proved generic uniqueness of pure-strategy Nash equilibrium, and uniqueness of the equilibrium outcome, for a class of non-atomic games where a player's payoff depends on, and strictly decreases with, the measure of the set of players playing the same (pure) strategy it is playing.
They also proved generic uniqueness of the Cournot--Nash equilibrium distribution, corresponding to a description of a game in terms of distribution of player types.

A crowding game is a game in which the payoff of each player depends only on the player's action and the size of the set of players choosing that particular action: the larger the set, the smaller the payoff.
\citet{milchtaich2000generic} proved that a large crowding game generically has just one equilibrium, and the equilibrium payoffs in such a game are always unique.
Moreover, the sets of equilibria of the \(m\)-replicas of a finite crowding game generically converge to a singleton as \(m\) tends to infinity.

\citet{milchtaich2005topological} proved topological conditions for uniqueness of equilibrium in physical networks with a large number of users (\emph{e.g.}, transportation, communication, and computer networks).

\citet{caines2013} and \citet{caines2018} surveyed \emph{Mean Field Game (MFG)} theory, which studies Nash equilibria in games involving a large number of agents.
They presented the main results of MFG theory, namely the existence and uniqueness of infinite-population Nash equilibria, their approximating finite-population \(\varepsilon\)-Nash equilibria, and the associated best-response strategies.

\subsection{Equilibrium Finding}

\citet{parise2019graphon} and \citet{parise2023graphon} presented a framework for analyzing equilibria and designing interventions for large network games sampled from a stochastic network formation process represented by a graphon.
Graphon games involve a continuum of agents whose payoff depends on their own action as well as a weighted average of other agents' actions, with heterogeneous weights specified by a graphon model.

\citet{perrin2020fictitious} analyzed continuous-time fictitious play on finite-state MFGs with additional common noise.
They presented a convergence analysis and proved that the induced exploitability decreases at a linear rate.
\citet{muller2022learning} introduced mean-field PSRO, an adaptation of \emph{Policy-Space Response Oracles (PSRO)}~\citep{Lanctot17:Unified} that learns Nash, coarse-correlated, and correlated equilibria in MFGs.
\citet{perolat2022scaling} tackled equilibrium computation in MFGs by using \emph{Online Mirror Descent (OMD)}. 
They proved that continuous-time OMD converges to a Nash equilibrium under certain conditions.
\citet{wang2023empirical} tackled MFGs using a version of the double oracle algorithm, iteratively adding (approximate) best responses to the equilibrium of the empirical MFG obtained from the strategies considered so far.
\citet{wu2024population} proposed a deep reinforcement learning algorithm that achieves population-dependent Nash equilibrium in MFGs without the need for averaging or sampling from history, inspired by Munchausen reinforcement learning~\citep{vieillard2020munchausen} and OMD.

\section{Problem Formulation}
\label{sec:formulation}

Throughout the paper, we use the following notation.
If \(\mu\) is a family of measures indexed by \(\mathcal{I}\), \({\bigotimes} \mu = \bigotimes_{i \in \mathcal{I}} \mu(i)\) is their product measure.
If \(f : \mathcal{X} \to \mathcal{Y}\), \(x \in \mathcal{X}\), and \(y \in \mathcal{Y}\), then
\(f[x \to y](t) = y\) if \(t = x\) and \(f(t)\) otherwise.
If \(\mathbf{x} \in \mathbb{R}^d\), \(\|\mathbf{x}\|\) is its Euclidean norm.
Let \(\mathbb{B}_d\) be the \(d\)-dimensional unit ball.
Let \(\overline{\mathbb{R}} = \mathbb{R} \cup \{-\infty, \infty\}\) be the order-completion of the real numbers, \emph{i.e.}, the extended real number line.
We now define several key concepts from game theory.

A \emph{game} is a tuple \((\mathcal{I}, \mathcal{S}, u)\) where
\(\mathcal{I}\) is the set of \emph{players},
\(\mathcal{S}(i)\) is the set of \emph{strategies} for player \(i\),
and \(u(s, i) \in \mathbb{R}\) is the \emph{utility} of player \(i\) under strategy profile \(s\).
A \emph{strategy profile} is an assignment of a strategy to each player, \emph{i.e.}, a function \(s\) for which \(s(i) \in \mathcal{S}(i)\) for each player \(i\).

A \emph{best response (BR)} for player \(i\) to a strategy profile \(s\) is a strategy that maximizes its utility given the other players' strategies, \emph{i.e.}, an element of \(\mathcal{B}(s, i) = \argmax_{x \in \mathcal{S}(i)} u(s[i \to x], i)\).

A \emph{Nash equilibrium (NE)} is a strategy profile \(s\) for which each player's strategy is a BR to the other players' strategies, \emph{i.e.}, \(s(i) \in \mathcal{B}(s, i)\) for all \(i \in \mathcal{I}\).
Nash equilibrium is the most common solution concept in game theory.

Let \(\mu\) be a measure on \(\mathcal{I}\).\footnote{
For example,
if \(\mathcal{I}\) is finite, this could be the counting measure, which simply yields the cardinality of the set.
If \(\mathcal{I} = [0, 1]^n\) for some \(n \in \mathbb{N}\), this could be the Lebesgue measure, which yields the ``volume'' of a subset of Euclidean space.
}
A \(\mu\)-Nash equilibrium, or \(\text{NE}_\mu\) for short, is a strategy profile \(s\) for which \(s(i) \in \mathcal{B}(s, i)\) for almost all \(i \in \mathcal{I}\).\footnote{
If \((\mathcal{X}, \Sigma, \mu)\) is a measure space, a property \(P\) holds \(\mu\)-almost everywhere if and only if the set of elements that do not satisfy \(P\) is contained in a \(\mu\)-null set, \emph{i.e.}, a measurable set \(\mathcal{A} \in \Sigma\) for which \(\mu(\mathcal{A})\) = 0.}
If every inhabited set has positive measure\footnote{
This is true, for example, for the counting measure.
}, \(\text{NE}_\mu = \text{NE}\).

The \emph{regret} of player \(i\) under strategy profile \(s\) is the highest utility it could gain from unilaterally changing its strategy, \emph{i.e.}, \(r(s, i) = \sup_{x \in \mathcal{S}(i)} u(s[i \to x], i) - u(s, i)\).
Thus \(r(s, i) \geq 0\).
Furthermore, \(s \in \text{NE}\) if and only if \(r(s, i) = 0\) for all \(i \in \mathcal{I}\).
Likewise, \(s \in \text{NE}_\mu\) if and only if \(r(s, i) = 0\) for almost all \(i \in \mathcal{I}\).

Let \(R(s) = \sup_{i \in \mathcal{I}} r(s, i)\).
Thus \(R(s) \geq 0\).
Furthermore, \(R(s) = 0\) if and only if \(s \in \text{NE}\).
Consequently, \(R\) is used as a standard metric of ``closeness'' to NE in the literature~\citep{Lanctot17:Unified, lockhart2019computing, walton2021multiagent, timbers2022approximate}.
The measure-aware equivalent \(R_\mu(s) = \esssup_{i \in \mathcal{I}} r(s, i)\) replaces the supremum with an essential supremum\footnote{
The essential supremum of a function \(f : \mathcal{X} \to \mathbb{R}\) is the infimum \(a \in \overline{\mathbb{R}}\) such that \(f(x) \leq a\) almost everywhere.
}, and \(R_\mu(s) = 0\) if and only if \(s \in \text{NE}_\mu\).

Let \(E(s) = \int_{i \sim \mu} r(s, i)\).
Since the integrand is non-negative, \(E(s) \geq 0\).
Furthermore, since the integrand is non-negative, if the integral is zero, the integrand is zero almost everywhere.
The converse is also true.
Thus \(E(s) = 0\) if and only if \(s \in \text{NE}_\mu\).
\(E\) is often called the ``exploitability'' or ``NashConv'' in the literature~\citep{Lanctot17:Unified, lockhart2019computing, walton2021multiagent, timbers2022approximate}.
Like \(R\), \(E\) is used as a standard metric of ``closeness'' to NE.
Indeed, when \(\mu\) is the counting measure, \(R\) and \(E\) can be bounded in terms of each other: \(E(s) / \mu(\mathcal{I}) \leq R(s) \leq E(s) \leq R(s) \mu(\mathcal{I})\).
The first and third inequalities hold in general, but not the second.
When \(\mu\) is a probability measure, \(E\) can be called the \emph{mean regret}.

\paragraph{Mixed strategies.}
Let \((\mathcal{I}, \mathcal{S}, u)\) be a game and, for each \(i \in \mathcal{I}\), let \(\Sigma(i)\) be a sigma-algebra on \(\mathcal{S}(i)\).
Then there exists a mixed-strategy game \((\mathcal{I}, \mathcal{S}', u')\) where, for each player \(i\), \(\mathcal{S}'(i)\) is the set of probability measures on \(\mathcal{S}(i)\) defined on \(\Sigma(i)\), and \(u'(s', i) = \expect_{s \sim {\bigotimes} s'} u(s, i)\).
That is, each player's strategy is a probability measure over its original strategy set; such a strategy is called a \emph{mixed strategy}.
Furthermore, its utility is the expected utility in the original game.
A \emph{mixed-strategy NE (MSNE)} of the original game is an NE of this mixed-strategy game.

\paragraph{Continuous-action game.}
A continuous-action game is a game whose strategy sets are subsets of Euclidean space, \emph{e.g.}, \(\mathcal{S}(i) \subseteq \mathbb{R}^d\).
The following theorems apply to such games.
\citet{Nash50:Equilibrium} showed that if \(\mathcal{S}(i)\) is nonempty and finite, a mixed-strategy NE exists.
\citet{Glicksberg52:Further} showed that if \(\mathcal{S}(i)\) is nonempty and compact, and \(u(s, i)\) is continuous in \(s\), a mixed-strategy NE exists.
\citet{Glicksberg52:Further,fan1952fixed,debreu1952social} showed that if \(\mathcal{S}(i)\) is nonempty, compact, and convex, and \(u(s, i)\) is continuous in \(s\) and quasiconcave in \(s(i)\), a pure strategy NE exists.
\citet{Dasgupta86:Existence} showed that if \(\mathcal{S}(i)\) is nonempty, compact, and convex, and \(u(s, i)\) is upper semicontinuous and graph continuous in \(s\) and quasiconcave in \(s(i)\), a pure strategy NE exists.
They also showed that if \(\mathcal{S}(i)\) is nonempty, compact, and convex, and \(u(s, i)\) is bounded and continuous in \(s\) except on a subset (defined by technical conditions) and weakly lower semicontinuous in \(s(i)\), and \(\sum_{i \in \mathcal{I}} u(s, i)\) is upper semicontinuous in \(s\), a mixed-strategy NE exists.
\citet{rosen1965existence} proved the uniqueness of a pure NE for continuous-action games under diagonal strict concavity assumptions.

\section{Optimization Techniques}
\label{sec:optimization}

In this section, we review the optimization techniques that we leverage.
We phrase them for the context of our problem.

\paragraph{Simultaneous gradient ascent.}
Let \((\mathcal{I}, \mathcal{S}, u)\) be a game.
For any strategy profile \(s\) and player \(i\), let
\begin{align} \label{eq:sim_grad}
    v(s, i) = \frac{\diff u(s, i)}{\diff s(i)}
\end{align}
This is called the \emph{simultaneous gradient (SG)}.
It consists of the derivative of each player's utility with respect to \emph{its own strategy}, as if the other players' strategies are fixed.

A standard method for tackling multiplayer games is \emph{simultaneous gradient ascent (SGA)}. 
It consists of discretizing the ordinary differential equation \(\dot{s} = v(s)\) in time---that is, letting \(s_{t+1} = s_t + \alpha_t v(s_t)\) where \(s_t\) and \(\alpha_t > 0\) are the strategy profile and stepsize at iteration \(t\), respectively.
In other words, on each iteration, each individual player tries to incrementally improve its own utility, as if the other players were fixed.\footnote{
SGA can be considered a differential or infinitesimal version of \emph{best-response dynamics}, since on each iteration each player \(i\) jumps to a best response to the other players' strategies under a local linear approximation to \(i\)'s utility function plus a quadratic distance penalty.
This is analogous to how standard gradient ascent jumps to the maximum of a quadratically-regularized local linear approximation to the loss function.
That is, \(x_{t+1} = \argmax_x f'(x_t) (x - x_t) - \frac{1}{2 \alpha_t} \|x - x_t\|_2^2 = x_t + \alpha_t f'(x_t)\).
}

The conditions under which SGA converges to a Nash equilibrium have been analyzed in many works.
\citet{mertikopoulos2019learning} prove that, if the game admits a pseudoconcave potential or if it is monotone, the players' actions converge to Nash equilibrium, no matter the level of uncertainty affecting the players' feedback.
\citet{bichler2021learning} note that most auctions in the literature assume symmetric bidders and symmetric equilibrium bid functions~\cite{Krishna02:Auction}; this symmetry creates a potential game, and SGA provably converges to a pure local Nash equilibria in finite-dimensional continuous potential games~\citep{mazumdar2020gradient}.
Thus in any symmetric, smooth auction game, symmetric gradient ascent with appropriate (square-summable but not summable) step sizes almost surely converges to a local ex-ante approximate Bayes-Nash equilibrium~\cite[Proposition 1]{bichler2021learning}.

\paragraph{Optimistic gradient ascent.}
A related method is \emph{optimistic gradient ascent (OGA)}.
It iterates \(s_{t+1} = s_t + \alpha_t v(s_t) + \beta_t (v(s_t) - v(s_{t-1}))\), where \(\alpha_t, \beta_t > 0\).\footnote{
The standard version of OGA uses \(\alpha_t = \beta_t\).
}
That is, it adds to SGA an extra term proportional to the difference of the current and past SG.
Intuitively, it uses this term to create an extrapolation (\emph{a.k.a}, prediction) of the future SG \(v(s_{t+1})\), and updates according to this prediction instead of the current SG \(v(s_t)\).
OGA converges in some games where SGA fails to converge.
OGA has been analyzed by \citet{popov1980modification}, \citet{daskalakis2017training}, and \citet{hsieh2019convergence}, among others.

There are also other learning dynamics in the literature, which have been surveyed and analyzed by \citet{balduzzi2018mechanics}, \citet{letcher2018stable}, \citet{letcher2019differentiable}, \citet{mertikopoulos2019learning}, \citet{mazumdar2019finding}, \citet{hsieh2021limits}, and \citet{willi2022cola}, among others, and these could be used as the optimizer within our proposed method just as well.

\paragraph{Simultaneous pseudo-gradient.}
Both SGA and OGA---as well as other learning dynamics in the literature---require computing the SG, which in turn requires computing derivatives of utilities with respect to strategies.
However, in some situations, the SG does not exist because the utility function is not differentiable.
In other situations, the utility function \emph{is} differentiable, but obtaining an unbiased estimator of its gradient is difficult or intractable.
This can happen if, for example, the utility is an expectation of a non-differentiable payoff function.
An example of such a situation is an auction, wherein a player wins an item as soon as its bid exceeds a threshold, yielding a payoff discontinuity.
In such a situation, it might be the case that the \emph{ex ante} utilities are differentiable but the \emph{ex post} utilities are not~\citep{bichler2021learning}.

To resolve this problem, we can replace each gradient in the definition of the SG with a \emph{pseudo-gradient}, which is an unbiased estimator of the gradient of a smoothed version of the function being differentiated.
The smoothed function is the convolution of the original function with a Gaussian kernel, which is equivalent to perturbing inputs to the original function with samples from that Gaussian and then taking an expectation.
Formally, if \(f : \mathbb{R}^d \to \mathbb{R}\) and \(\mu\) is the \(d\)-dimensional standard normal distribution, we have
\begin{align}
    \nabla_\mathbf{x} \expect_{\mathbf{z} \sim \mu} f(\mathbf{x} + \sigma \mathbf{z}) = \expect_{\mathbf{z} \sim \mu} \tfrac{1}{\sigma} f(\mathbf{x} + \sigma \mathbf{z}) \mathbf{z}.
\end{align}
Such pseudo-gradients are used in the zeroth-order optimization literature to optimize functions with access to only function values and not gradients.
Examples include \citet{sun2009stochastic}, \citet{wierstra2014natural}, \citet{duchi2015optimal}, \citet{nesterov2017random}, \citet{shamir2017optimal}, \citet{salimans2017evolution}, \citet{lenc2019non}, and \citet{berahas2022theoretical}.

In our case, we can use the estimator
\begin{align}
    \tilde{v}(s, i) = \tfrac{1}{\sigma} u(s[i \to s(i) + \sigma z(i)], i) z(i)
\end{align}
where \(z\) is a sample from the standard normal distribution of same shape as \(s\).
This is the approach presented by \citet{bichler2021learning}.
It requires one perturbation for each player and subsequent evaluation of the utility function.
\citet{martin2024joint} presented a way to reduce the number of utility function evaluations required per iteration from linear to constant in the number of players, based on the so-called \emph{pseudo-Jacobian}
\begin{align}
    \tilde{v}(s, \cdot) = \tfrac{1}{\sigma} u(s + \sigma z, \cdot) \odot z
\end{align}
where \(\odot\) is the Hadamard (\emph{a.k.a.}, elementwise) product.
This version only requires a single perturbation of the strategy profile and subsequent evaluation of the utility function, assuming that the evaluation yields utilities for all players simultaneously (as is often the case for game implementations in practice).

\paragraph{Mixed strategies over continuous action spaces.}
\citet{martin2023finding} studied the problem of computing an approximate Nash equilibrium of continuous-action game without access to gradients.
They model players' strategies as artificial neural networks.
In particular, they use \emph{randomized policy networks} to model mixed strategies.
These take noise in addition to an observation as input and can represent arbitrary observation-dependent, continuous-action distributions.
Being able to model such mixed strategies is crucial for tackling continuous-action games that lack pure-strategy equilibria.
We use this technique of input noise injection in games where we may need mixed strategies, that is, where we may need the ability to randomize over a continuous pure strategy space.

\section{Proposed Method}
\label{sec:method}

In this section, we present our method.

\paragraph{Player-to-Strategy Network.}
Suppose we are interested in tackling a game with infinitely many players.
This raises the problem of how to represent a strategy profile that has an infinite number of players.
We propose a way to do this which we coin a \emph{Player-to-Strategy Network (P2SN)}.
This is a neural network that takes as input a \emph{player} and outputs a \emph{strategy} for that player.
Neural networks are a universal class of function approximators and have a powerful ability to generalize well across inputs~\citep{cybenko1989approximation,hornik1989multilayer,hornik1991approximation,leshno1993multilayer,pinkus1999approximation}.
We exploit the strong generalization capabilities of neural networks to represent (and potentially learn, as we will see shortly) across an infinite number of possible inputs, \emph{i.e.}, players, simultaneously.\footnote{P2SN could be useful even in games with a finite number of players.
For instance, the input could be a one-hot encoding of the player's index, along with any other \emph{a priori} features of that player.
It allows for neural net parameters to be shared across players, while simultaneously taking into account each player's individual and specific \emph{a priori} features as input.
This can potentially result in faster learning and better generalization.
However, the present paper focuses on games with an infinite number of players.}

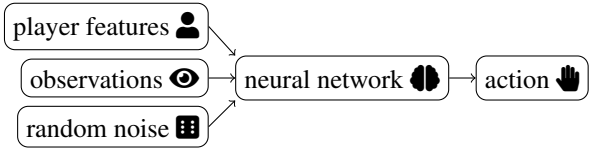
\begin{figure}
\centering
\begin{tikzpicture}[
node distance=0.2em and 1em,
every node/.style={draw, rounded corners},
tips=proper
]
\node (network) {neural network \faBrain};
\node (player) [above left=of network] {player features \faUser};
\node (observation) [left=of network] {observations \faEye};
\node (noise) [below left=of network] {random noise \faDiceSix};
\node (action) [right=of network] {action \faHandPaper};
\draw [->]
    (player.east) edge (network.north west)
    (observation.east) edge (network.west)
    (noise.east) edge (network.south west)
    (network.east) edge (action.west)
;
\end{tikzpicture}
\caption{High-level structure of the P2SN.}
\label{fig:p2sn}
\end{figure}

What specifically goes into the network as input depends on the game.
The inputs can include
(1) the features of the player (\emph{e.g.}, their X and Y coordinates in a spatial game),
(2) any observations a player receives during the game\footnote{
The games we test on do not have such additional observations, but others might.
},
and (3) random noise that allows the network to randomize over actions (in order to generate mixed strategies).
This is shown in Figure~\ref{fig:p2sn}.

\paragraph{Fourier features.}
The space of player feature vectors \(\mathcal{I}\) is often low-dimensional (e.g., it would be 2-dimensional in the location game on a square), and it is important for our P2SN to be able to represent fine details on such a space.
It has been found that standard feedforward neural networks struggle to learn detailed, high-frequency features in low-dimensional input spaces.
\citet{rahaman2019spectral} showed that standard neural networks are biased towards low-frequency functions, meaning that they cannot have local fluctuations without affecting their global behavior.
More precisely, using Fourier analysis, they showed that these networks prioritize leaning the low-frequency modes, a phenomenon they call \emph{spectral bias}.
\citet{tancik2020fourier} showed that passing input points through a simple \emph{Fourier feature mapping} enables a \emph{multilayer perceptron (MLP)} to learn high-frequency functions in low-dimensional problem domains, whereas a standard MLP has impractically slow convergence to high frequency signal components.

To let our P2SN more easily represent and learn fine details on a low-dimensional player feature space, which we found to be crucial in some experiments, we preprocess player features to the network with the map \(F : \mathbf{i} \mapsto \sin(\mathbf{B} \mathbf{i})\), where \(\mathbf{i}\) is the input vector of player features and \(\mathbf{B} \in \mathbb{R}^{d \times N}\) is a learnable matrix, where \(d\) is the dimensionality of the player feature space and \(N \gg d\) is a large number.
In our experiments, we use \(N = 64\).
We initialize the entries of \(\mathbf{B}\) with independent samples from the normal distribution with standard deviation \(\sigma = 64\).
Thus \(F\) represents a learnable mapping of the input player to a high-dimensional space of Fourier (\emph{a.k.a.}, frequency-domain) features.\footnote{
If the observation space is also low-dimensional, we could also apply a learnable Fourier feature mapping to it before passing it to the rest of the network.
The need for that did not arise in the experiments in this paper.
}
In our implementation, we made the entries of \(\mathbf{B}\) be part of the neural network, so they are trained the same way as the rest of the network parameters.

\paragraph{Shared-Parameter Simultaneous Gradient (SPSG).}
Since we now share parameters between players, instead of maintaining individual disjoint parameters for each player, Equation~\ref{eq:sim_grad} no longer makes sense.
This is because it requires taking a derivative with respect to the parameters \emph{of a specific player and no other}, which is no longer possible.\footnote{
Consider what would happen in a two-player zero-sum game where both players share parameters.
The parameters would not change at all, because Equation~\ref{eq:sim_grad} would ``pull'' them in exactly equal and opposite directions.
}
Furthermore, since we are in an infinite-player setting, we would need to compute these derivatives for an \emph{infinite} number of players.

To solve these problems, we re-express the simultaneous gradient in a way that relies on taking the derivative \emph{with respect to the entire strategy profile all at once}:
\begin{align}
    v(s) = \mleft[ \frac{\diff}{\diff r} \int_{i \sim \mu} u(s[i \mapsto r(i)], i) \mright]_{r = s}
\end{align}
Here, \([\cdot]_{r = s}\) means ``evaluate the expression inside the brackets, treated as a function of \(r\), at the argument \(s\)''.
Also, the derivative \(\frac{\diff}{\diff r}\) of the integral is a \emph{functional derivative}, that is, the derivative of a functional (the integral) with respect to a function (namely \(r\)).

This generalization \emph{does} apply to the shared-parameter case, where \(s\) is the set of parameters for the P2SN.
Consequently, we call it the \emph{Shared-Parameter Simultaneous Gradient (SPSG)}.
In practice, SPSG can be implemented as follows.
First, let \(\hat{\mu} = \frac{\mu}{\mu(\mathcal{I})}\) be the normalized probability measure corresponding to \(\mu\).\footnote{
We assume that \(0 < \mu(\mathcal{I}) < \infty\), \emph{i.e.}, \(\mu\) is normalizable.
}
On each iteration:
\begin{enumerate}
    \item Create a copy \(r\) of the current strategy profile \(s\).
    \item Sample a player \(i\) from \(\hat{\mu}\).
    \item Create a hybrid strategy profile \(s[i \mapsto r(i)]\) that intertwines \(s\) and \(r\), using \(r\)'s output for player \(i\) and \(s\)'s output for any other player.
    \item Pass this strategy profile to the utility function and get player \(i\)'s utility.
    \item Multiply player \(i\)'s utility by \(\mu(\mathcal{I})\).
    \item Take the gradient of this scalar with respect to \(r\).\footnote{
Though \(r\) and \(s\) have equal \emph{values}, they are different \emph{variables} and may have different gradients.
In the last step, we take the gradient with respect to \(r\) specifically, not \(s\), which is used in a different way inside the expression \(s[i \mapsto r(i)]\).
To illustrate this, consider the function \(f(x, y) = x y^2\). At \(x = y = 1\), \(x\) and \(y\) have the same value.
However, the gradient with respect to \(x\) is 1, while the gradient with respect to \(y\) is 2.
    }
\end{enumerate}
The player-sampling procedure in Step (2) above allows us to obtain an unbiased estimator of the entire integral, which in turn allows us to obtain an unbiased estimator of the SPSG.
(The integral and derivative commute with each other.)
All we need for this procedure is an unbiased estimator of the utility function \(u(s, i)\).
Such an estimator can often be obtained for infinite-player games.
For example, many infinite-player games can be expressed as a series of the general form
\begin{align}
    u(s, i) &= f(i, s(i)) + \int_{j \sim \nu(i)} \Big( g(i, j, s(i), s(j)) \\
    & + \int_{k \sim \xi(i, j)} h(i, j, k, s(i), s(j), s(k)) + \ldots \Big)
\end{align}
where \(f\), \(g\), \(h\), \emph{etc.} are arbitrary functions and \(\nu(i)\), \(\xi(i, j)\), \emph{etc.} are arbitrary measures.
That is, player \(i\)'s utility is an integral over pairwise interactions with other players, 3-way interactions with other pairs of players, \emph{etc.}, up to some limited order.
In that case, an unbiased estimator of player \(i\)'s utility can be obtained by first sampling \(j\) given \(i\), then sampling \(k\) given \(i\) and \(j\), and so on, for however many terms are necessary.
If necessary, we can use techniques like \emph{Markov chain Monte Carlo (MCMC)}~\citep{metropolis1953equation,hastings1970monte} to obtain these samples.
As we will see in the experiments section, many games of interest require only the first integral, \emph{i.e.}, involve only the aggregation of pairwise interactions between players (and their corresponding strategies).
One special kind of game is an \emph{aggregative game}, which has a utility function of the form
\begin{align}
    u(s, i) &= f\mleft( i, s(i), \int_{j \sim \nu(i)} s(j) \mright)
\end{align}
That is, each player's utility depends on its own strategy and an aggregate of all players' strategies.
Two examples of such a game are the Ising game (\S\ref{sec:ising}) and Cournot competition (\S\ref{sec:cournot}).

We emphasize that our method makes no assumption of symmetry or identicality across players.
Each player can have its own arbitrary strategy space and utility function.
For example, in spatial games of \S\ref{sec:experiments}, the agents are clearly not identical: some are close to the boundary (where boundary effects come into play), while others are distant from it.

\section{Experiments}
\label{sec:experiments}

We evaluated the proposed techniques using computational experiments.
For each experiment, we ran 8 trials.
In each graph, solid lines show the mean across trials, and bands show the standard error of the mean.
In the neural net, on each iteration, we use a batch size of \(256\), averaging gradients across that batch.
For the P2SN, we use 2 hidden layers of size 64 each, the swish activation function~\citep{jahan2023self}, and He initialization~\citep{he2015delving} to initialize the weights.
Each experiment was run individually on one NVIDIA A100 SXM4 40GB GPU.
All experiments took between 9 and 14 minutes of training time.
For our experiments, we use
Python 3.12.3,
jax 0.4.34~\citep{jax2018github},
flax 0.8.5~\citep{flax2020github},
optax 0.2.4~\citep{deepmind2020jax},
matplotlib 3.9.2~\citep{Hunter:2007},
and scipy 1.14.1~\citep{2020SciPy-NMeth}.

The regret attained by our method depends on hyperparameters like learning rate and network size, which controls the flexibility of the strategy profile representation.
We use a small network because finding the best one is outside the scope of the paper and irrelevant to our goal.
Our goal is to present our method, which is to our knowledge the first of its kind, and show it can learn to approximate NE.

To estimate mean regret across players, we discretize the space of players into \(N\) points\footnote{
We discretize a 1-dimensional interval in the obvious way, \emph{i.e.}, by placing \(N\) equally-spaced points across the interval.
We discretize a higher-dimensional cube by taking the first \(N\) points of the Roberts sequence~\citep{roberts2018unreasonable}, a low-discrepancy sequence (\emph{a.k.a.} quasirandom sequence) with good approximation properties.
} and compute the regret of each of these individual players, followed by an average.
To estimate an individual player's regret, we discretize its strategy space into \(N\) points.
We then find the strategy with the highest expected utility and subtract the player's current expected utility.
Expected utilities are estimated by averaging \(N = 200\) samples from the stochastic utility function.

\subsection{Ising Games}
\label{sec:ising}

The first game in the experiments models a situation where there is a continuum of players, each of whom wants to satisfy a particular endogenous \emph{bias}, but also wants to exogenously \emph{conform} (or anti-conform) to its neighbors.
It is called ``the Ising game'' because it resembles the Lenz--Ising model~\citep{lenz1920beitrvsge,ernst1925beitrag} of statistical mechanics, which models ferromagnetism and the magnetic dipole moments of atomic spins.
In particular, each player's strategy corresponds to an \emph{expected spin vector}, and the integral of the utility function over all players (\emph{i.e.}, the utilitarian social welfare) yields the negative Hamiltonian of that model.
This game has been studied by \citet{galam2010ising}, \citet{xin2017ising}, \citet{leonidov2020qre}, \citet{leonidov2024ising}, and \citet{feldman2024sharp}, among others.

Formally, let \(\mathcal{I}\) be the space of possible player vectors.
Let \(\mathcal{S}(i) = \mathbb{B}_d\) be the set of strategies for player \(i\), where \(d \in \mathbb{N}\).
Let \(\mathbf{b} : \mathcal{I} \to \mathbb{R}^d\) be a \emph{bias field}.
Let \(\nu(i)\) be a measure on \(\mathcal{I}\).\footnote{In general, \(\nu(i)\) could be a \emph{signed} measure, which can take on negative values, which yields an incentive to \emph{anti}-conform to neighboring players.}
Let the utility for player \(i\) under strategy profile \(\mathbf{s}\) be
\begin{align}
    u(\mathbf{s}, i) &= \underbrace{\mathbf{s}(i) \cdot \mathbf{b}(i)}_\text{bias} + \underbrace{\int_{j \sim \nu(i)} \mathbf{s}(i) \cdot \mathbf{s}(j)}_\text{conformity}.
\end{align}

Every strategy set is convex and compact, and each player's utility is linear in its own strategy and affine in the other players' strategies.
Thus a pure-strategy Nash equilibrium \(\mathbf{s}^\ast\) exists.

First, we experimented with the 1D version of this game where \(d = 1\) and \(\mathcal{I} = [0, 1]\).
Let \(\nu(i)\) be a Gaussian measure of scale \(\sigma = \tfrac{1}{10}\) centered at \(i\) and truncated to \(\mathcal{I}\).\footnote{This means that all points outside \(\mathcal{I}\) get zero measure density, \emph{i.e.}, only points inside \(\mathcal{I}\) contribute to the integral. Formally, \(\operatorname{truncate}(\mu, \mathcal{I})(\mathcal{A}) = \mu(\mathcal{A} \cap \mathcal{I})\).}
Let\footnote{
The constants for this and other problems in this paper were chosen (before training) to ensure that the respective fields have enough peaks and troughs to make the problems interesting.
They are analogous to the arbitrary constants used in various test functions for optimization, such as Rosenbrock's function \citep{rosenbrock1960automatic}.
}
\begin{align}
    b(i) &= \sin(10 \pi i) + \cos(14 \pi i).
\end{align}

\begin{figure}
    \centering
    \includegraphics{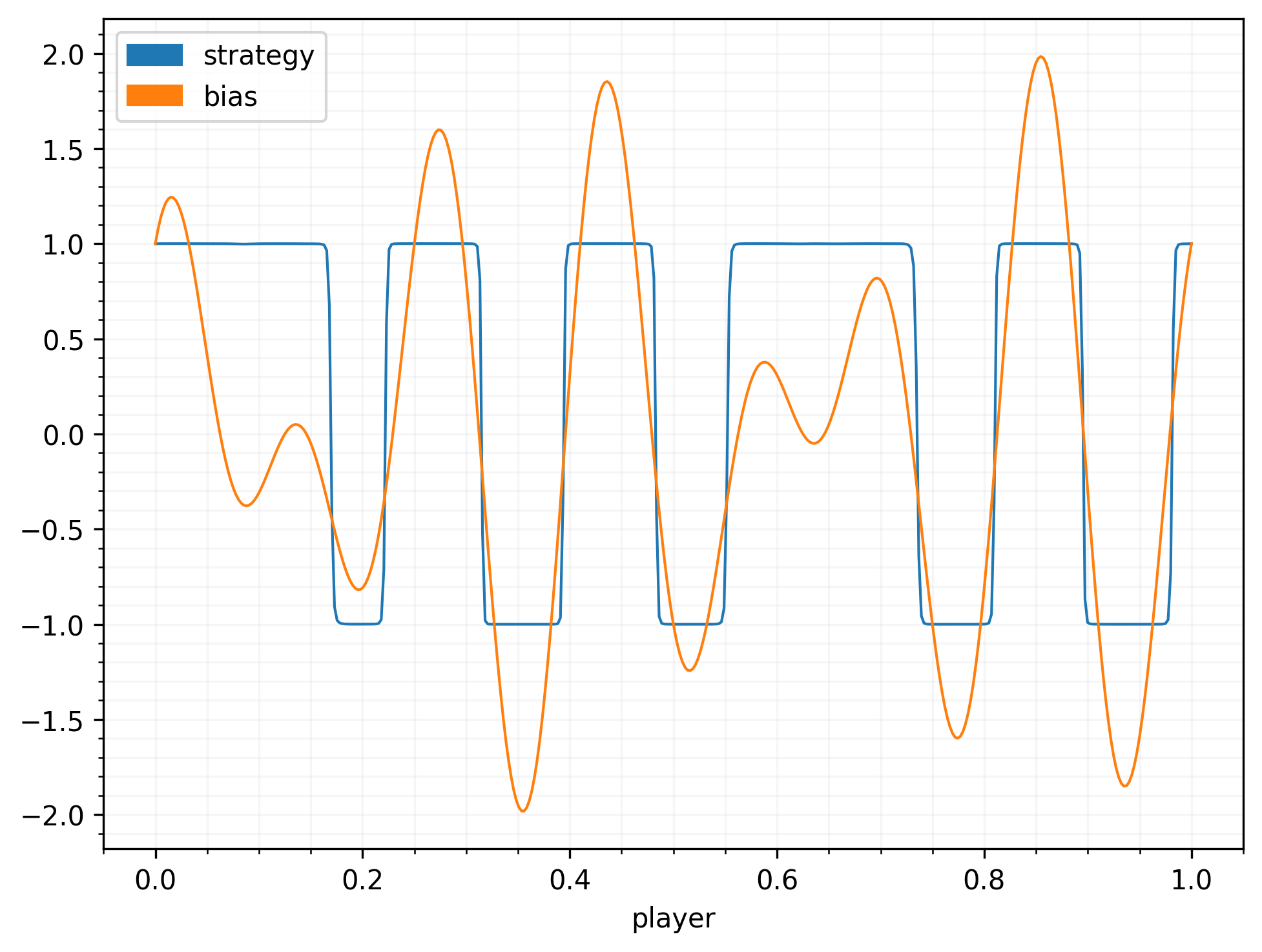}
    \caption{Strategy profile for 1D Ising game.}
    \label{fig:ising_1d_profile}
\end{figure}

\begin{figure}
    \centering
    \includegraphics{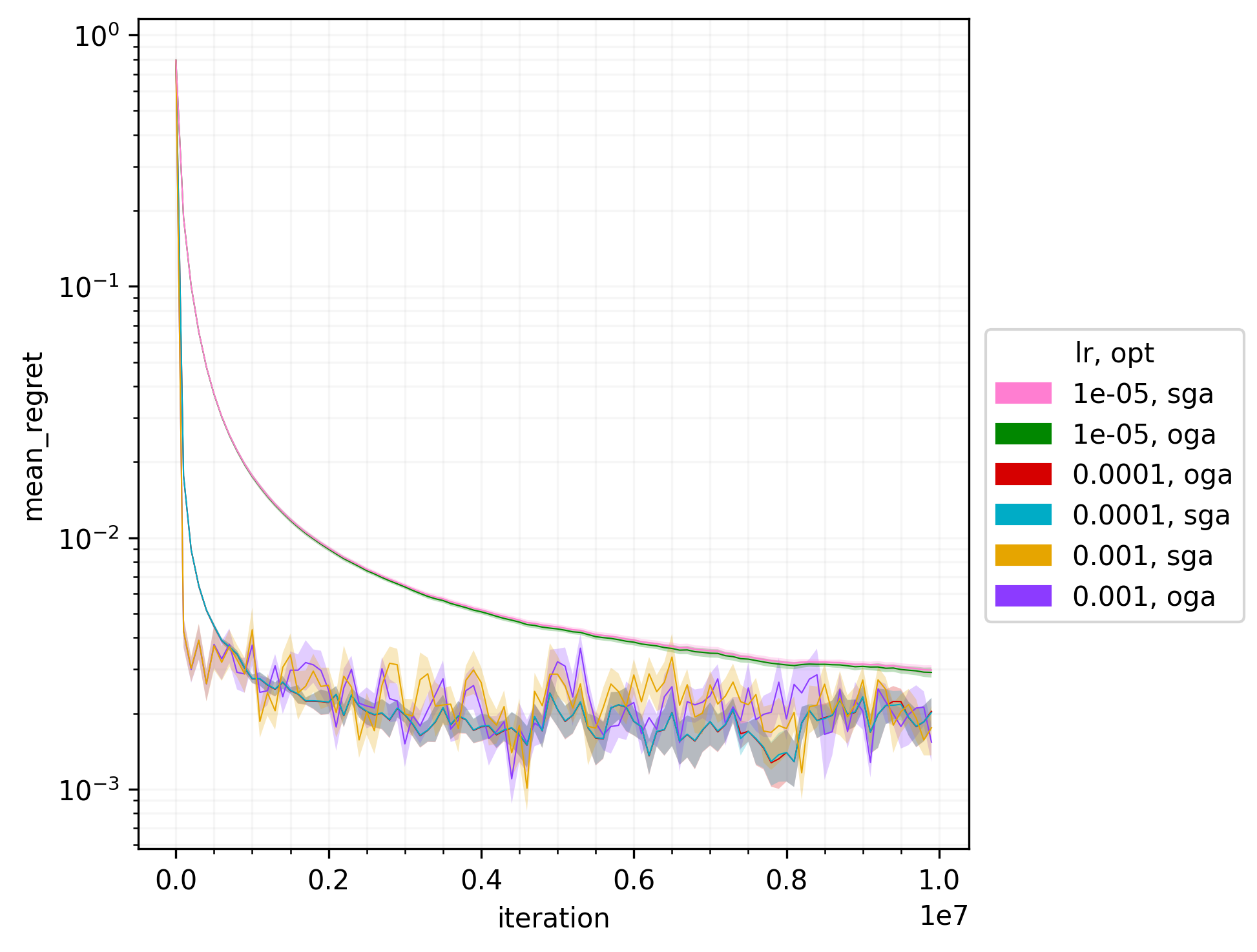}
    \caption{Regrets for 1D Ising game.}
    \label{fig:ising_1d_regret}
\end{figure}

The strategy profile learned by our algorithm is illustrated in Figure~\ref{fig:ising_1d_profile}.
The learned strategy profile switches sharply between \(-1\) and \(1\) multiple times across the unit interval of players.
Regrets over the course of training are shown in Figure~\ref{fig:ising_1d_regret}.
The regrets decrease toward zero over time as training proceeds.

\begin{figure}
    \centering
    \includegraphics{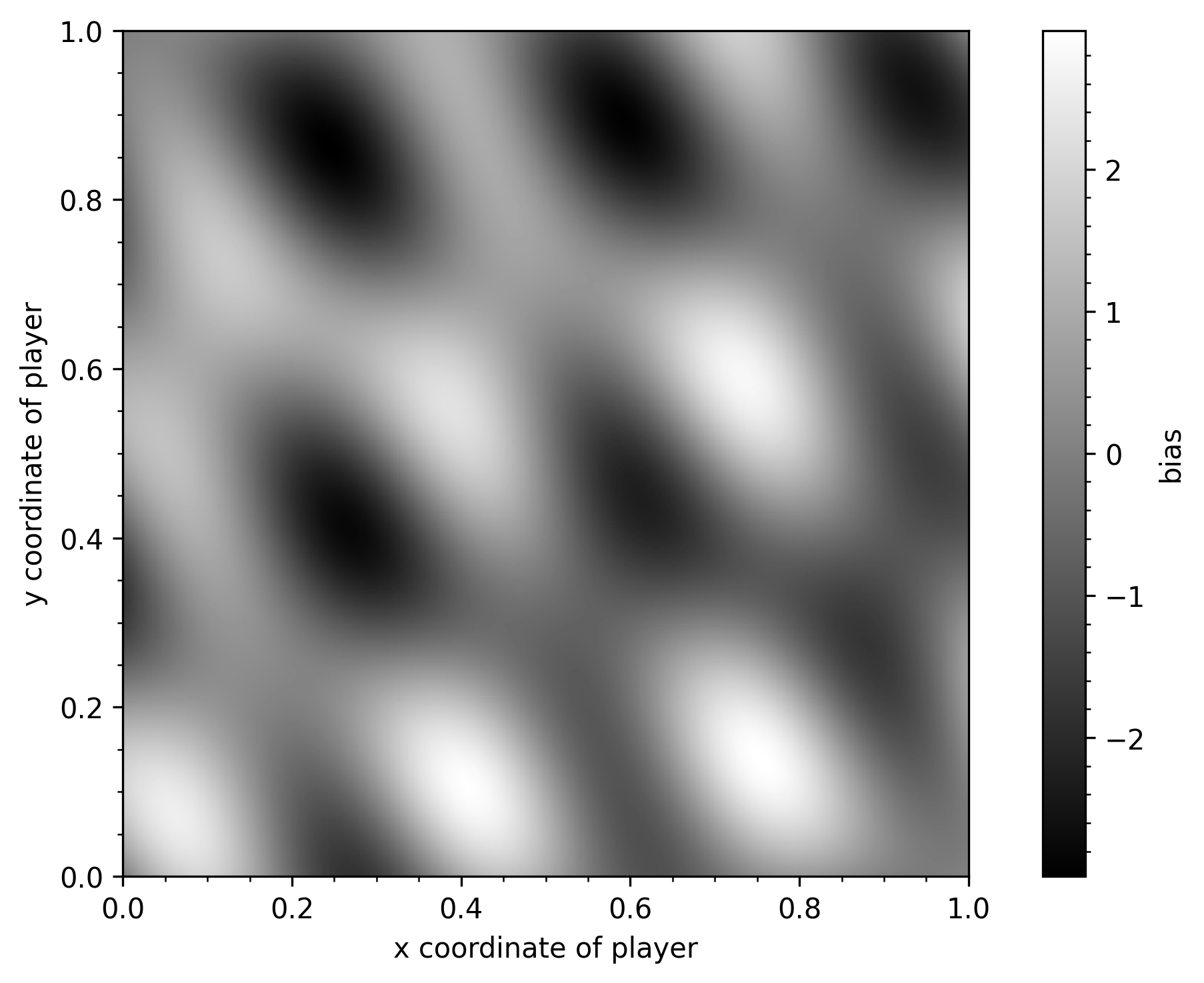}
    \caption{Bias field for 2D Ising game.}
    \label{fig:ising_2d_bias}
\end{figure}

\begin{figure}
    \centering
    \includegraphics{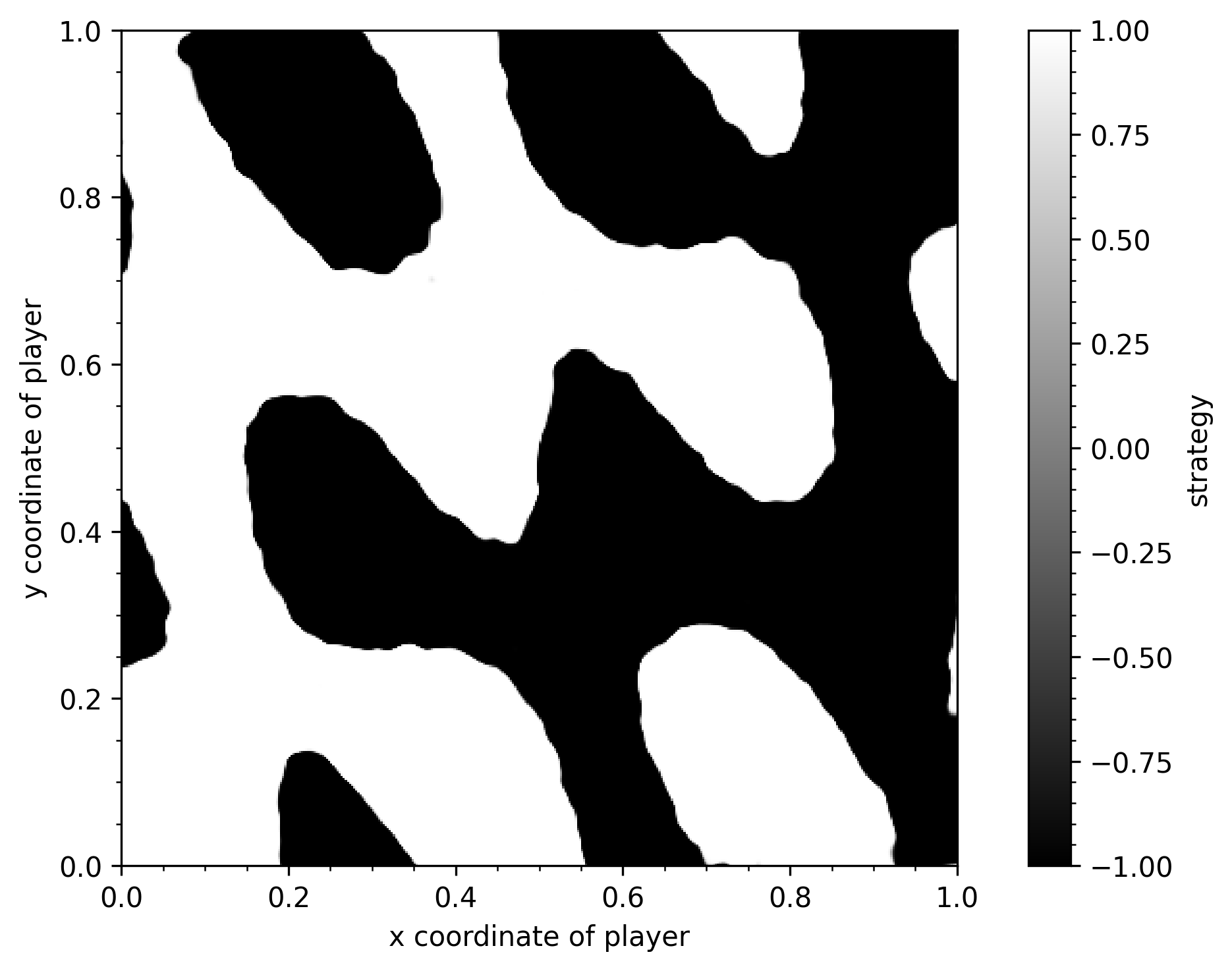}
    \caption{Strategy profile for 2D Ising game.}
    \label{fig:ising_2d_profile}
\end{figure}

\begin{figure}
    \centering
    \includegraphics{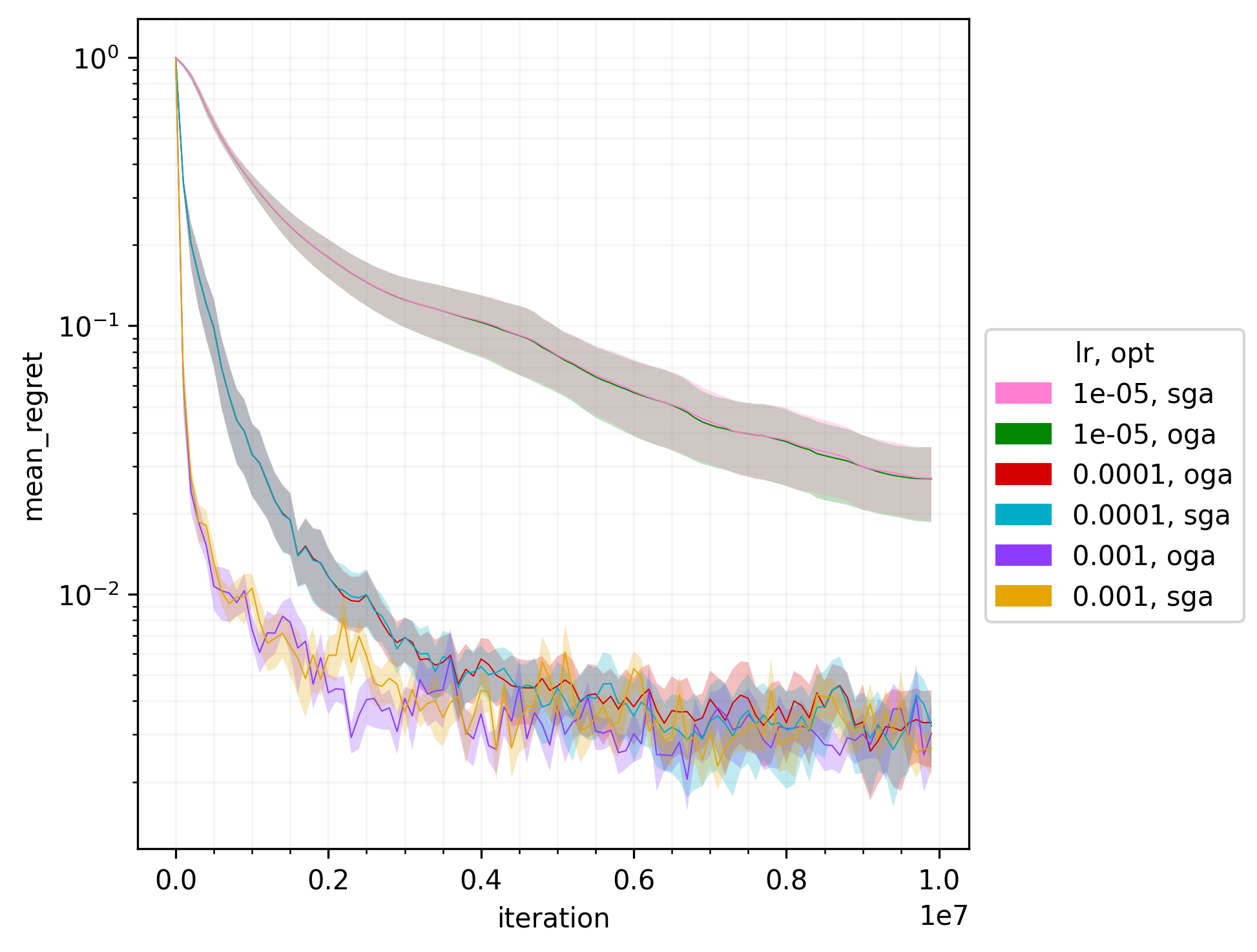}
    \caption{Regrets for 2D Ising game.}
    \label{fig:ising_2d_regret}
\end{figure}

Second, we experimented with the 2D version of this game where \(\mathcal{I} = [0, 1]^2\) and
\begin{align} \label{eq:ising_bias_2d}
    b(x, y) &= \sin(4 \pi x) + \sin(6 \pi y) + \sin(5 \pi (x + y)).
\end{align}
where \((x, y) = i\).
This field is illustrated in Figure~\ref{fig:ising_2d_bias}.
On this game, our algorithm learns the strategy profile shown in Figure~\ref{fig:ising_2d_profile}.
Regrets over the course of training are shown in Figure~\ref{fig:ising_2d_regret}.

\subsection{Distance-Based Ising Games}

Then, we experimented with the same games as before, but with a different utility function
\begin{align}
    u(\mathbf{s}, i)
    &= -\|\mathbf{s}(i) - \mathbf{b}(i)\|^2
    - \int_{j \sim \nu(i)} c(i, j) \|\mathbf{s}(i) - \mathbf{s}(j)\|^2.
\end{align}
Each player \(i\) picks a point \(\mathbf{s}(i)\) and seeks to minimize the Euclidean distance to a target point \(\mathbf{b}(i)\), but also to minimize the Euclidean distance to the points chosen by its neighbors.
Each player's utility is concave in its own strategy.

\begin{figure}
    \centering
    \includegraphics{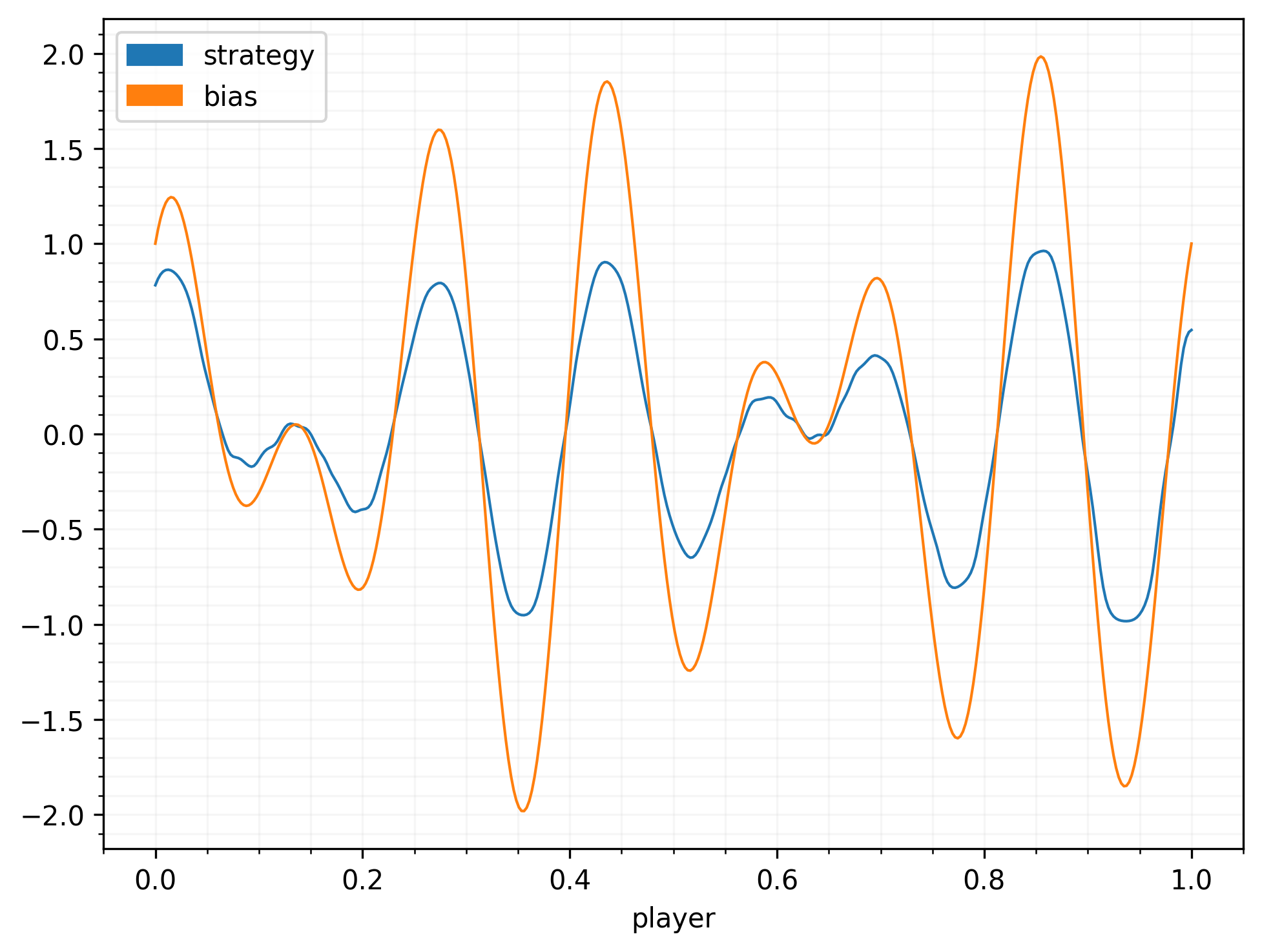}
    \caption{Strategy profile for 1D distance-based Ising game.}
    \label{fig:ising_distance_1d_profile}
\end{figure}

\begin{figure}
    \centering
    \includegraphics{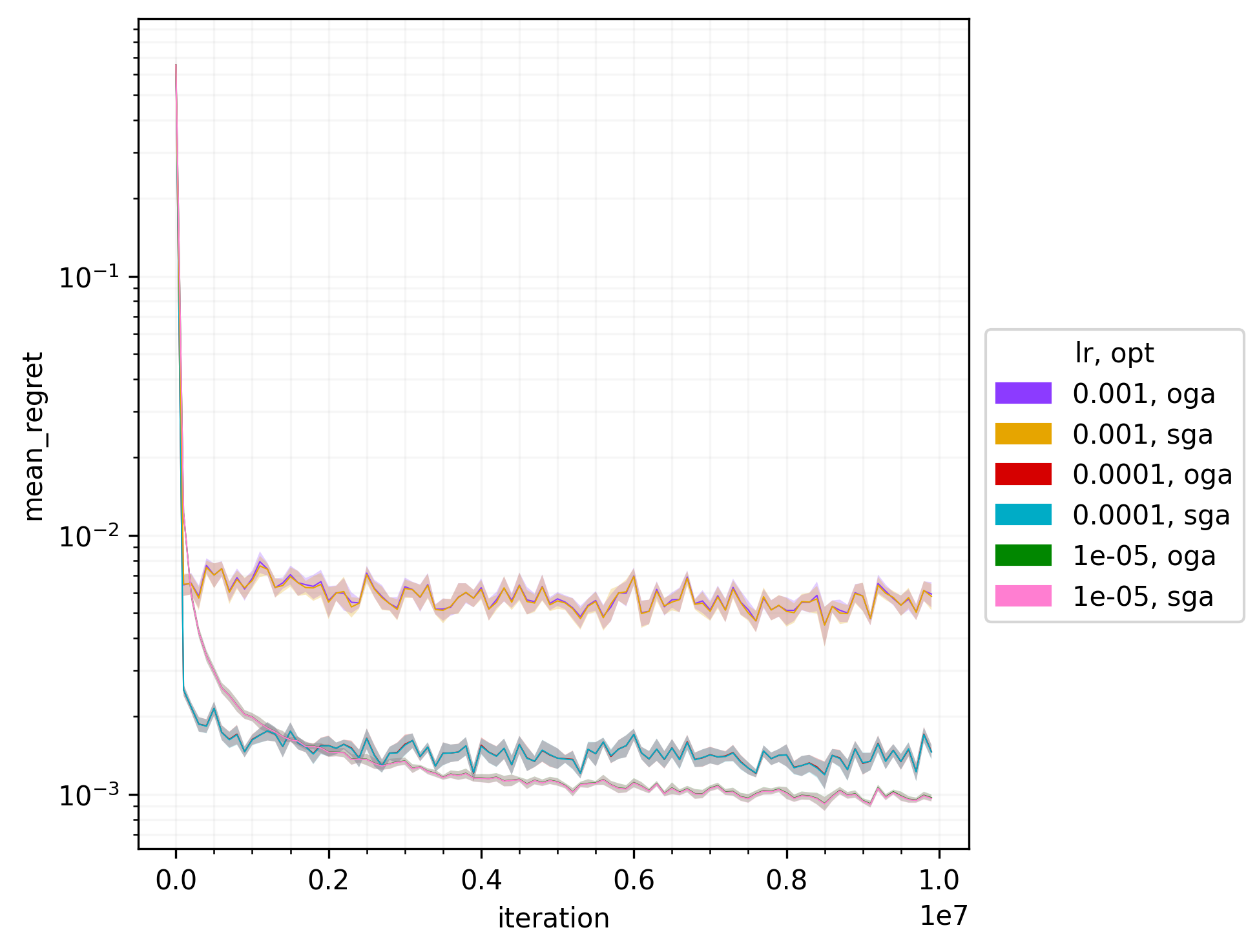}
    \caption{Regrets for 1D distance-based Ising game.}
    \label{fig:ising_distance_1d_regret}
\end{figure}

The strategy profile learned by our algorithm is illustrated in Figure~\ref{fig:ising_distance_1d_profile}.
Unlike the case with dot products, the strategies chosen by players now take on values that change continuously in the interior of the interval \([-1, 1]\).
Regrets are shown in Figure~\ref{fig:ising_distance_1d_regret}.
The regrets decrease toward zero over time as training proceeds.

\begin{figure}
    \centering
    \includegraphics{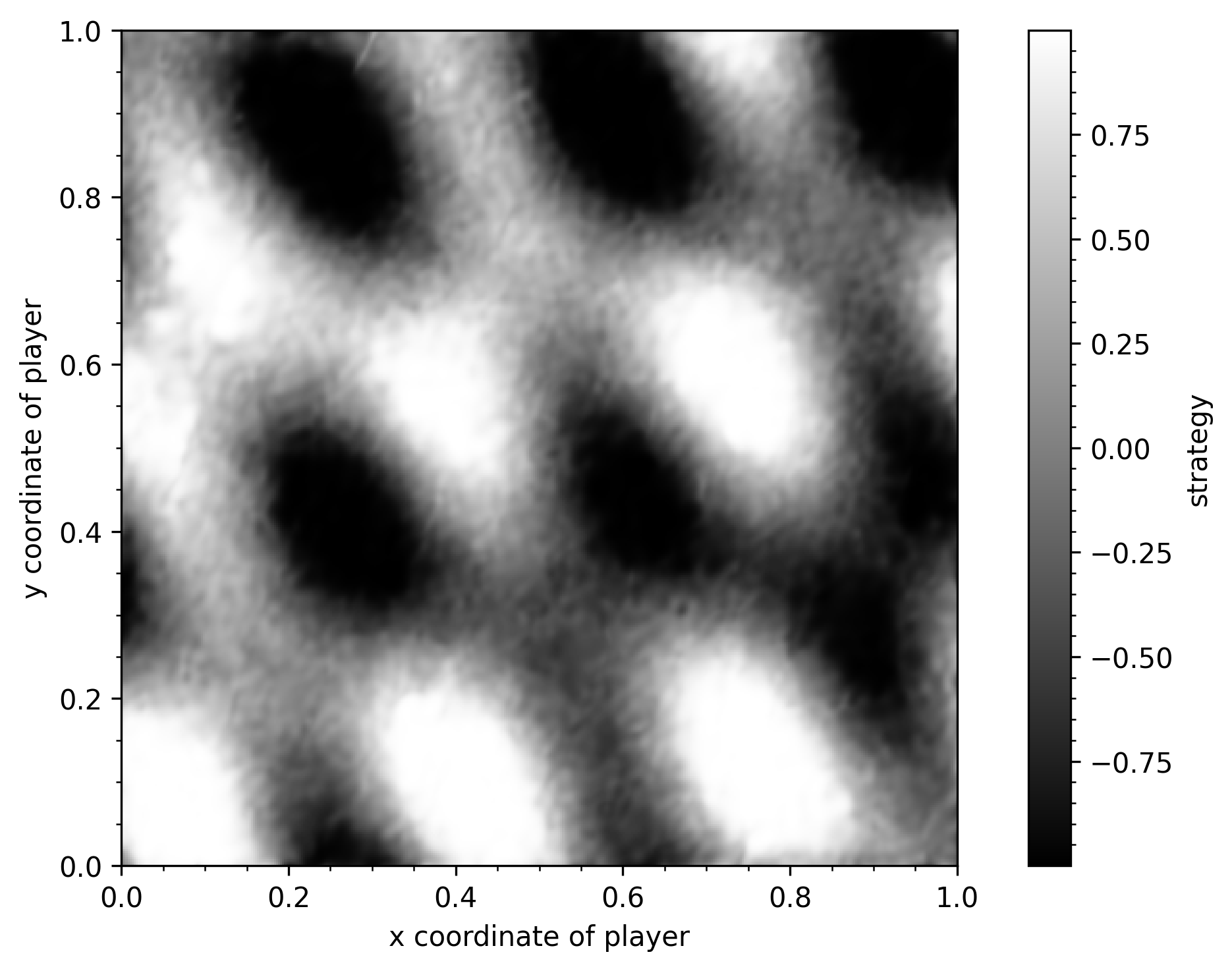}
    \caption{Strategy profile for 2D distance-based Ising game.}
    \label{fig:ising_distance_2d_profile}
\end{figure}

\begin{figure}
    \centering
    \includegraphics{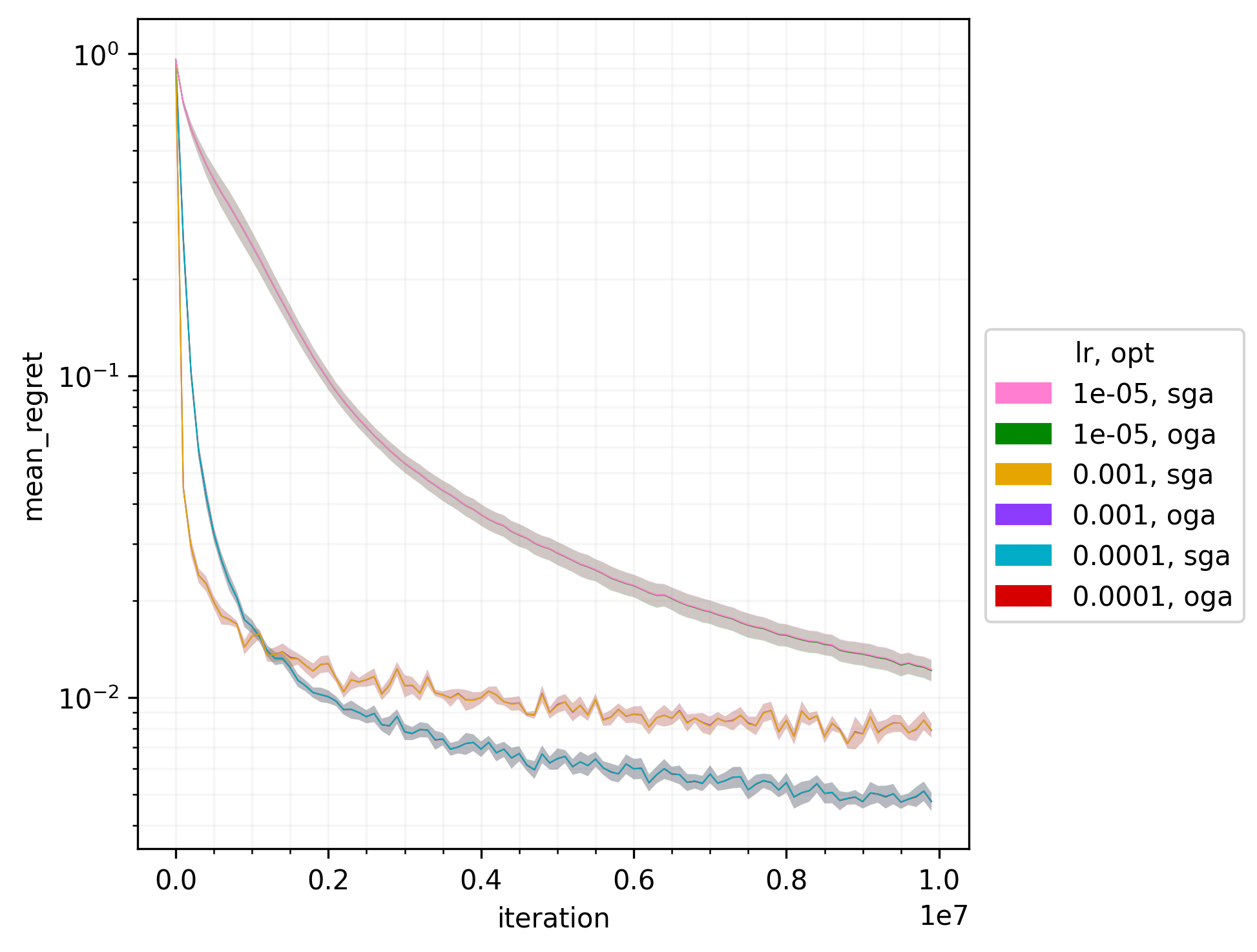}
    \caption{Regrets for 2D distance-based Ising game.}
    \label{fig:ising_distance_2d_regret}
\end{figure}

We also test on a 2-dimensional version with \(\mathcal{I} = [0, 1]^2\) and \(b\) as in Equation~\ref{eq:ising_bias_2d}, yielding the results shown in Figures~\ref{fig:ising_distance_2d_profile} and~\ref{fig:ising_distance_2d_regret}.

\subsection{Cournot Competition}
\label{sec:cournot}

Cournot competition is a classic economic model dating back to \citet{cournot1863principes} in which competing firms choose a quantity of output to produce independently and simultaneously.
Versions with a continuum of players have been studied extensively (\emph{e.g.},~\citep{novshek1985perfectly,chan2015bertrand}).

We formulate this game as follows.
Let \(\mathcal{I}\) be the set of firms.
Let \(\mathcal{S}(i) \subseteq \mathbb{R}_{\geq 0 }\) be the set of possible outputs for firm \(i\).
Firm \(i\)'s utility function is
\begin{align}
    u(q, i) &= \underbrace{q(i) P(Q)}_{\text{revenue}} - \underbrace{C(i, q(i))}_{\text{cost}}
\end{align}
where
\(\mu\) is a measure on \(\mathcal{I}\)\footnote{
For example, \(\mathcal{I}\) could be a finite set and \(\mu\) could be the counting measure.
},
\(Q = \int_{i \sim \mu} q(i)\) is the \emph{aggregate output},
\(P\) is the \emph{inverse demand function},\footnote{
This function yields the market price associated with an aggregate output.
It is determined by consumers' demand.
Typically, it is assumed to be anti-monotone, \emph{i.e.}, the higher the aggregate output, the lower the market price.
}
and \(C(i, \cdot)\) is the \emph{production cost function} of firm \(i\).

\begin{figure}
    \centering
    \includegraphics{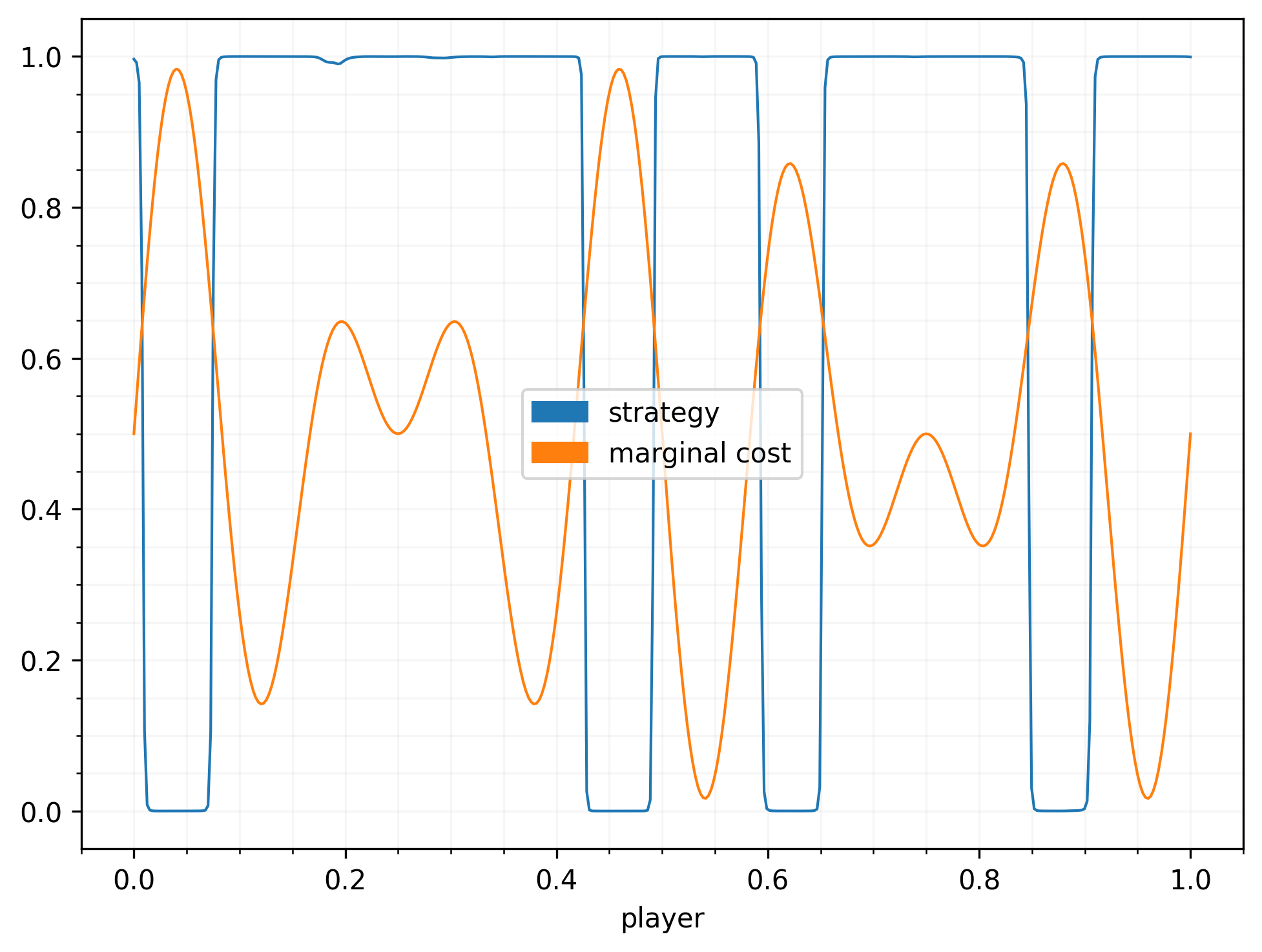}
    \caption{Strategy profile for Cournot competition.}
    \label{fig:cournot_profile}
\end{figure}

\begin{figure}
    \centering
    \includegraphics{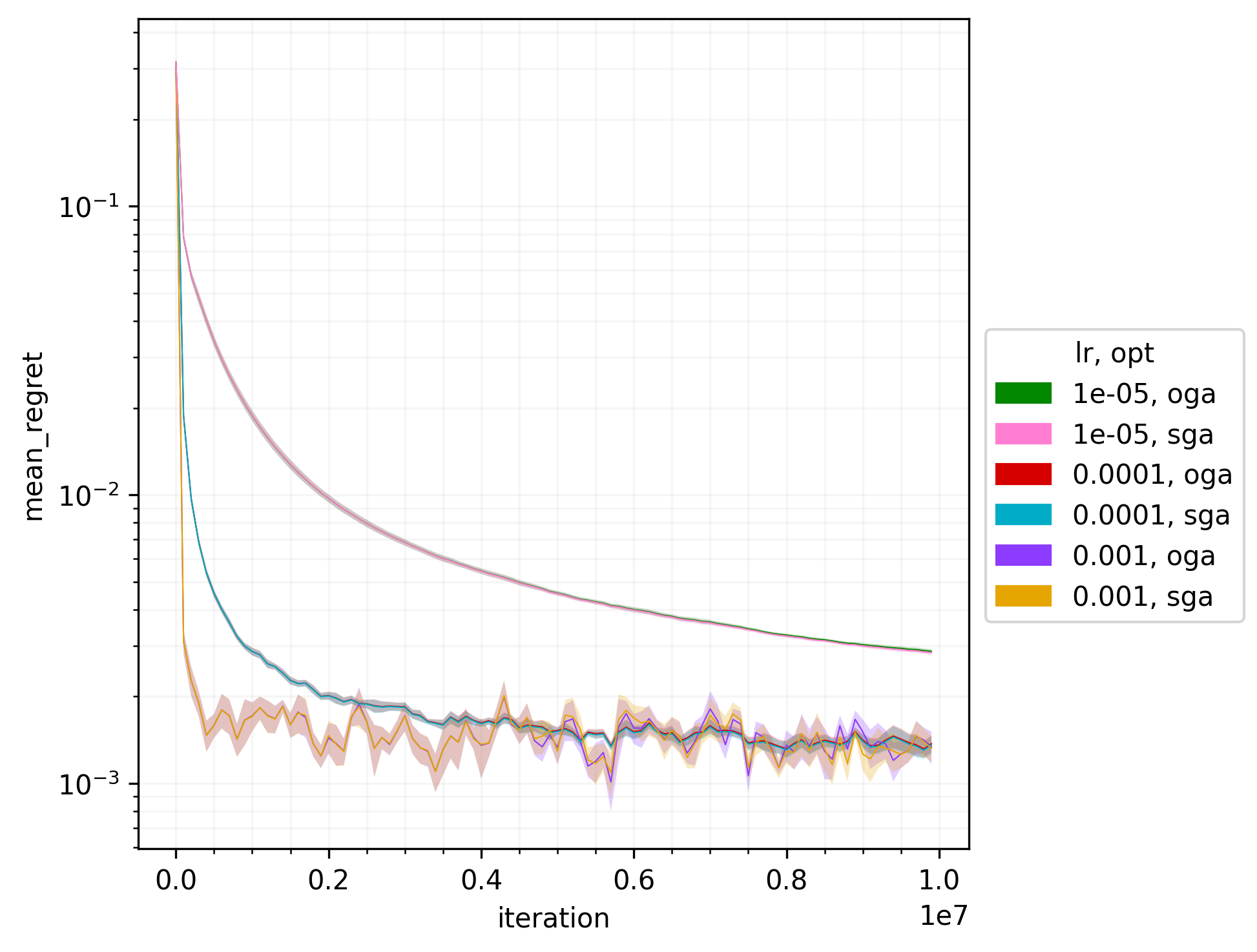}
    \caption{Regrets for Cournot competition.}
    \label{fig:cournot_regret}
\end{figure}

Suppose
\(P(Q) = a - b Q\) for some \(a, b \in \mathbb{R}\).
Also, suppose
\(C(i, x) = x c(i)\),
where 
\(c(i) \in \mathbb{R}\) is the \emph{marginal cost of production} for firm \(i\).
Now, consider the special case where
\(\mathcal{I} = [0, 1]\),
\(\mathcal{S}(i) = [0, 1]\),
\(\mu\) is the Lebesgue measure,
\(a = 2\),
\(b = 1.8\),
and
\begin{align}
    c(i) &= \tfrac{1}{2} +  \tfrac{1}{4} \sin(10 \pi i) + \tfrac{1}{4} \sin(14 \pi i).
\end{align}
The strategy profile learned by our algorithm is shown in Figure~\ref{fig:cournot_profile}.
The regrets of our algorithm over the course of training are shown in Figure~\ref{fig:cournot_regret}.
The regrets decrease toward zero over time as training proceeds.

\subsection{Local Cournot Competition}

We also experimented with a version of Cournot competition where the inverse demand function is player-specific and \emph{local}---that is, depends on a \emph{local} aggregate of outputs of nearby players, as opposed to a \emph{global} aggregate across all players.
Locality can be used to model spatial factors such as transportation costs or delivery time constraints.
Concretely, instead of using \(Q = \int_{j \sim \mu} q(j)\) to determine the global market price, we determine a local market price using a player-specific \(Q(i) = \int_{j \sim \nu(i)} q(j)\), where \(\nu(i)\) is some player-specific measure over \(\mathcal{I}\).

\begin{figure}
    \centering
    \includegraphics{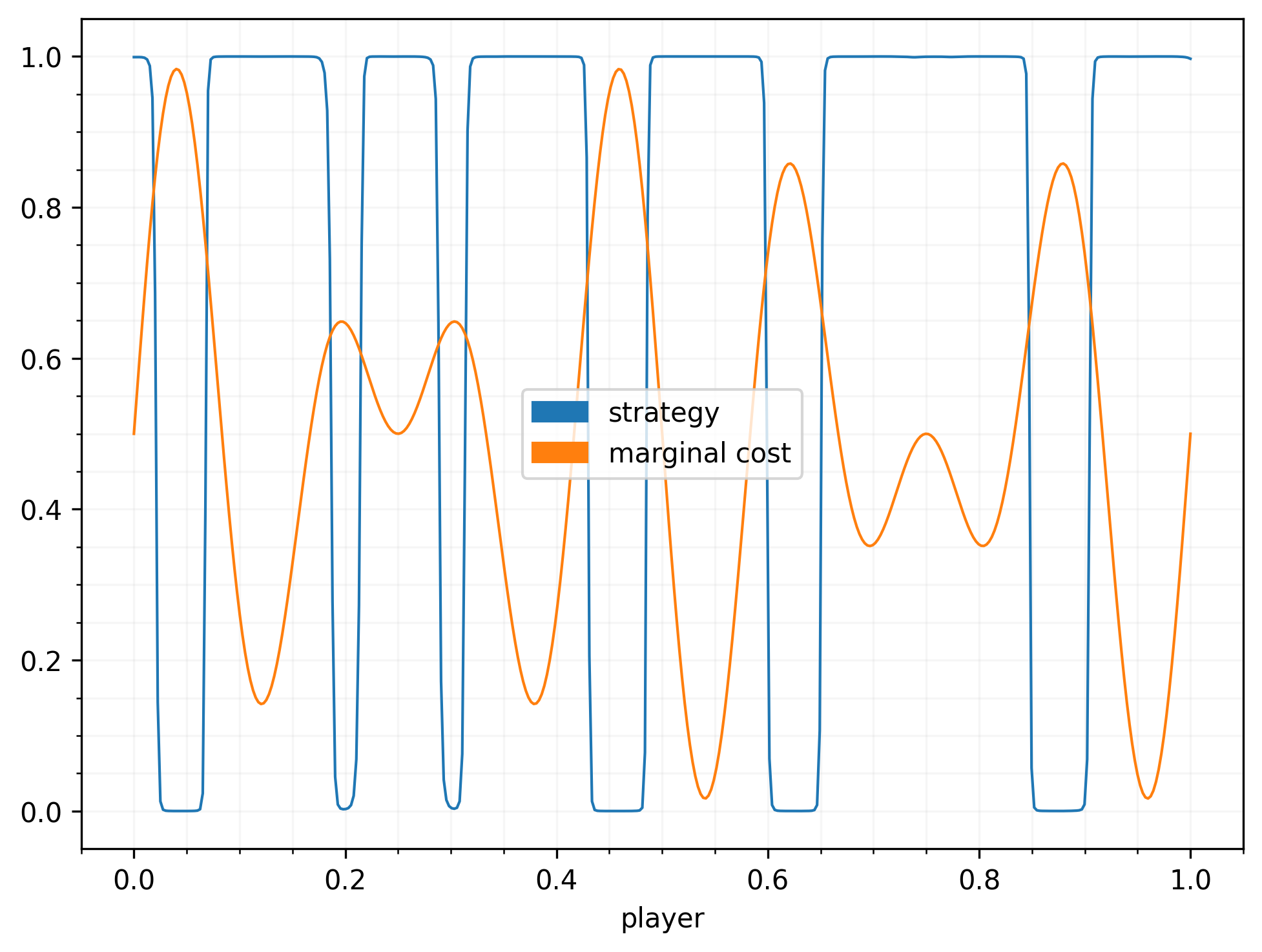}
    \caption{Strategy profile for local Cournot competition.}
    \label{fig:local_cournot_profile}
\end{figure}

\begin{figure}
    \centering
    \includegraphics{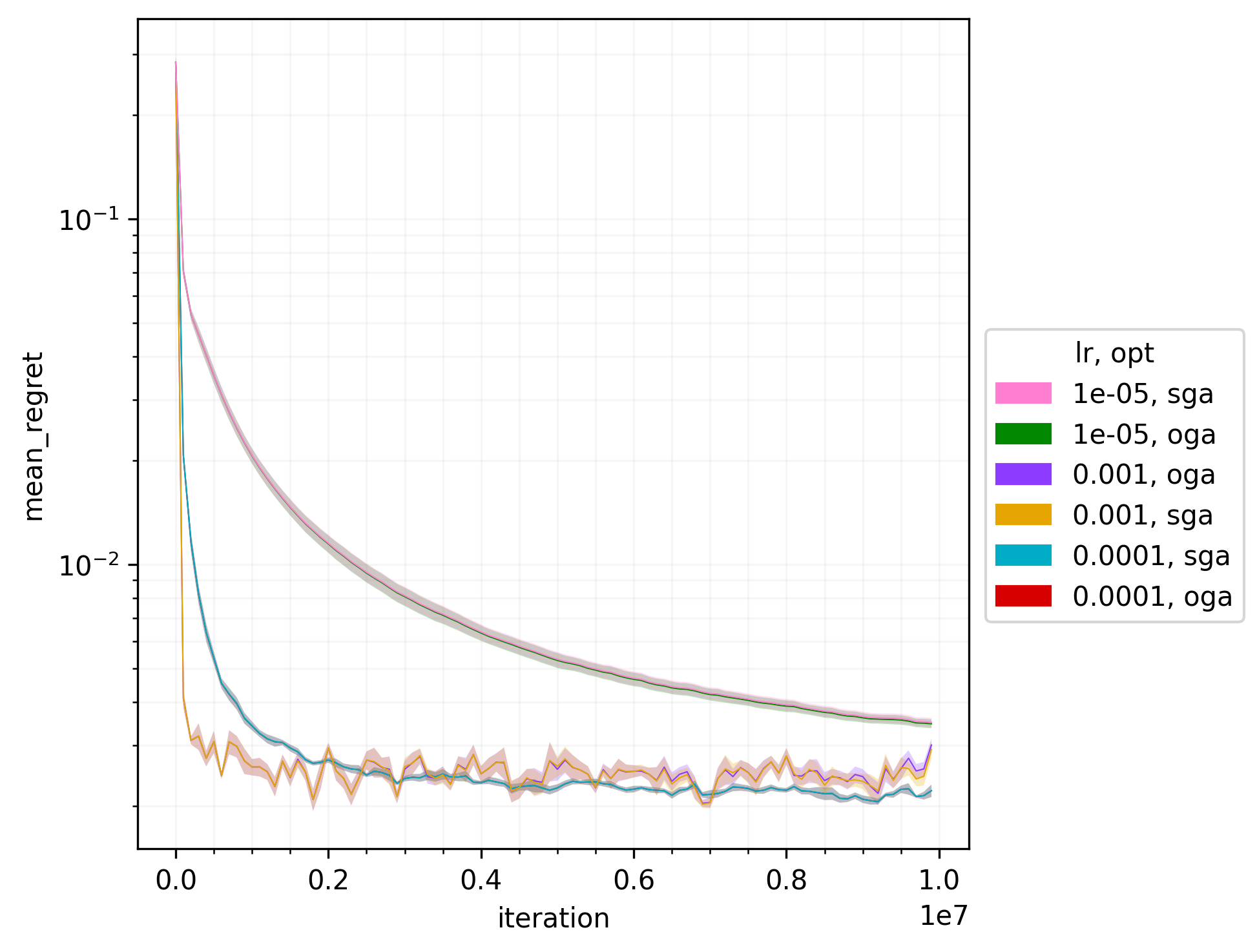}
    \caption{Regrets for local Cournot competition.}
    \label{fig:local_cournot_regret}
\end{figure}

We keep the same parameters as the original Cournot competition, including the marginal cost function, but use the local aggregation described above.
Specifically, we let \(\nu(i)\) be the Gaussian measure of scale \(\sigma = \tfrac{1}{10}\) centered at \(i\) and truncated to \(\mathcal{I}\).
The strategy profile learned by our algorithm is shown in Figure~\ref{fig:local_cournot_profile}.
The regrets of our algorithm over the course of training are shown in Figure~\ref{fig:local_cournot_regret}.
The regrets decrease toward zero over time.

\subsection{Crowding Game}

In this section, we experiment with an important subclass of congestion games called \emph{crowding games}.
Let \(d \in \mathbb{N}\), \(\mathcal{S}(i) = [0, 1]^d\), and
\begin{align}
    u(\mathbf{s}, i) = v(\mathbf{s}(i)) - \int_{j \sim \mu} K(\mathbf{s}(i), \mathbf{s}(j))
\end{align}
where \(v : \mathcal{S}(i) \to \mathbb{R}\) is a \emph{value function} and \(K\) is the \(d\)-dimensional Gaussian kernel of scale \(\sigma > 0\).
This game models a situation where players want to choose points that have high value, but that are also not too close to other players' points.
This kind of scenario can be found in many real-world settings, such as the spatial arrangement of crowds, buildings, infrastructure, and wildlife~\citep{carmona2020dyson}.

\begin{figure}
    \centering
    \includegraphics{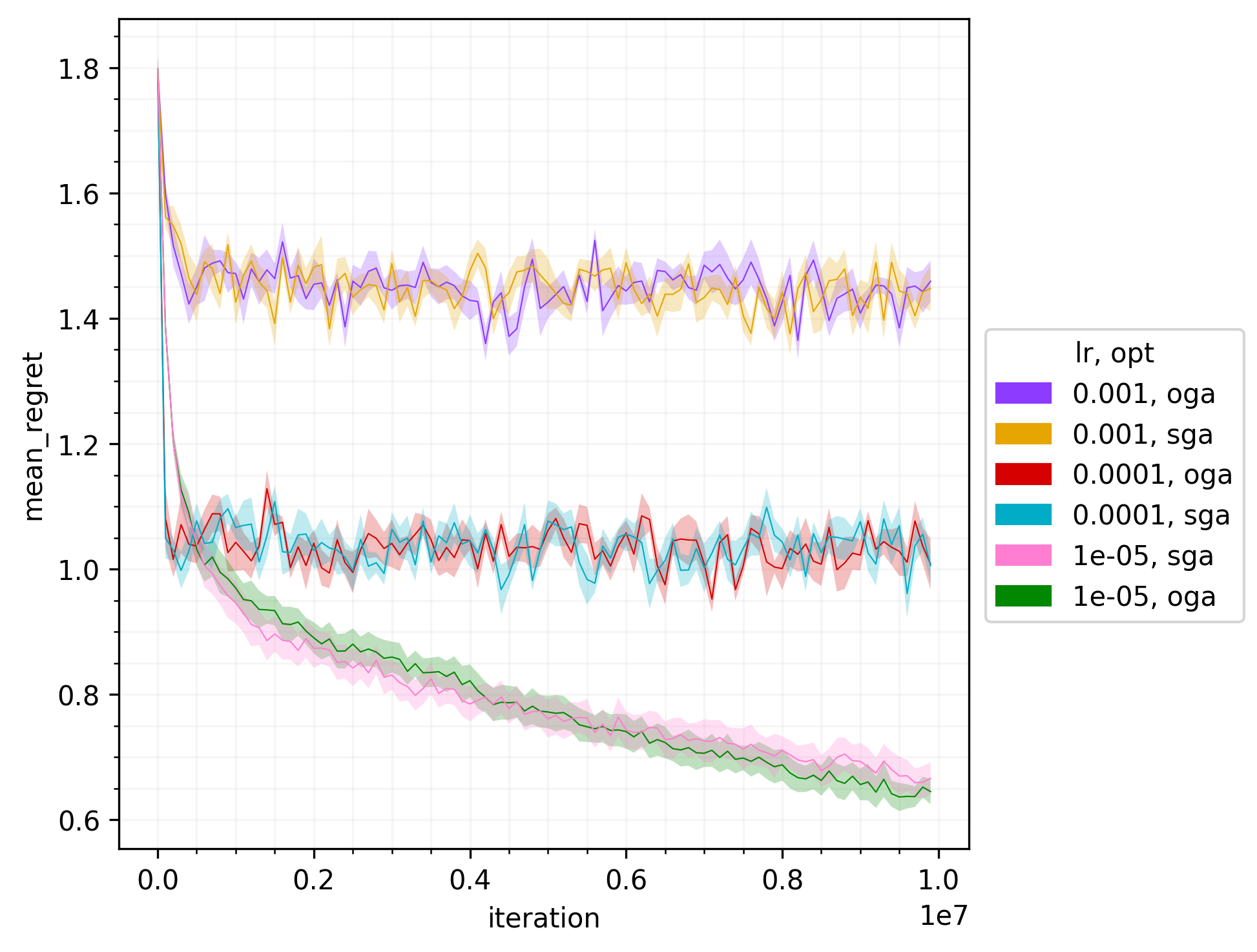}
    \caption{Regrets for crowding game.}
    \label{fig:crowding_regret}
\end{figure}

\begin{figure}
    \centering
    \includegraphics{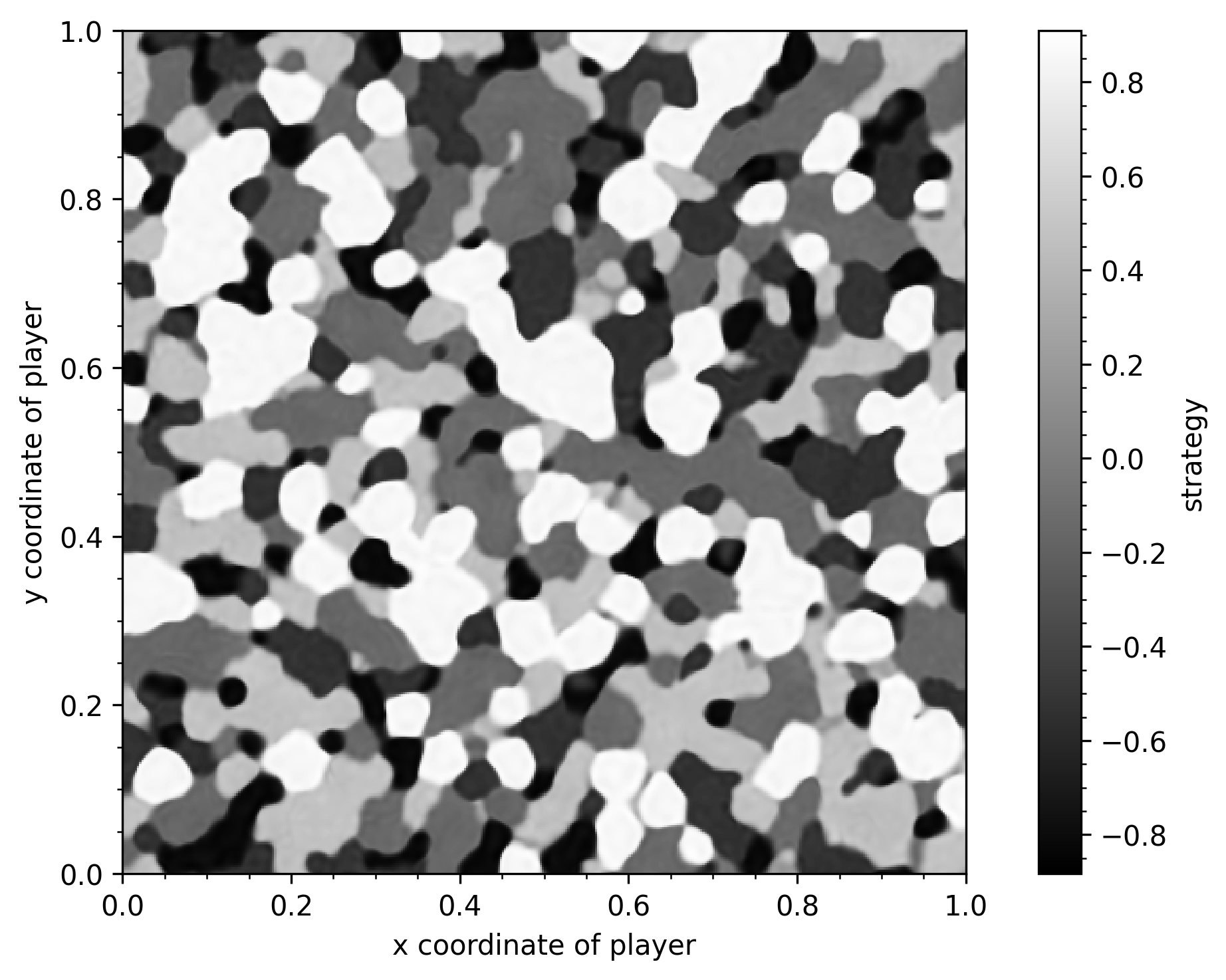}
    \caption{Learned strategy profile for crowding game.}
    \label{fig:crowding_profile}
\end{figure}

\begin{figure}
    \centering
    \includegraphics{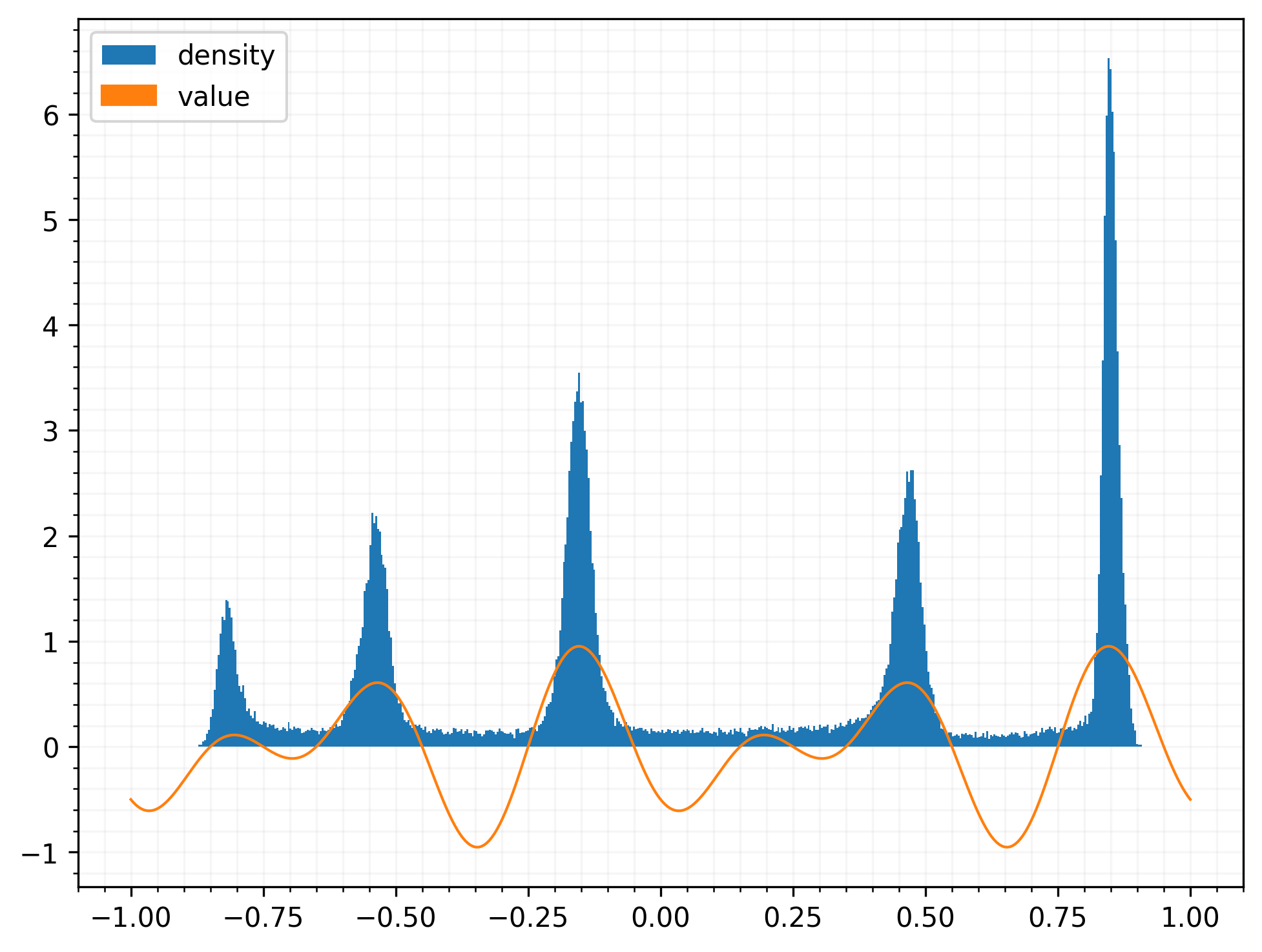}
    \caption{Histogram of strategies for crowding game.}
    \label{fig:crowding_histogram}
\end{figure}

We run an experiment with \(d = 2\), \(\sigma = 0.01\), and
\begin{align}
    v(s) = -\tfrac{1}{2} \sin(4 \pi s) -\tfrac{1}{2} \cos(6 \pi s)
\end{align}
Regrets are shown in Figure~\ref{fig:crowding_regret}.
An example of a learned strategy profile is shown in Figure~\ref{fig:crowding_profile}.
The corresponding histogram of strategies themselves is shown in Figure~\ref{fig:crowding_histogram}.
As expected, the strategies tend to concentrate around areas of high value.

\section{Conclusions and Future Research}
\label{sec:conclusion}

We presented a new method for solving games with a (countably or uncountably) infinitely number of players.
It uses a \emph{Player-to-Strategy Network (P2SN)} to represent a strategy profile for an infinite number of players, and trains it using \emph{Shared-Parameter Simultaneous Gradient (SPSG)}, an extension of classical simultaneous gradient ascent.
We tested our method on multiple infinite-player games from the literature and observed that it converges to an approximate Nash equilibrium in these games.

Our method generalizes simultaneous gradient ascent and its variants, which are classical equilibrium-seeking dynamics used for multiagent reinforcement learning.
Our method is capable in principle of handling games with infinitely many states, infinitely many players, infinitely many actions (and mixed strategies on them), and discontinuous utility functions.
One direction for future research is to tackle multi-step and even continuous-time infinite-player games. 
Another is to explore how broadly the convergence of our method can be theoretically proven.

As described in \S\ref{sec:related}, many games have a unique NE.
However, some games could have multiple NE.
NE refinements and equilibrium selection are outside the scope of this paper, but potentially an interesting question for future research.
Even without such future study, the presented techniques can be very useful. In many-player games, game-theoretic approaches have been successful and often the most successful. 
For example, Pluribus \citep{Brown19:Superhuman} for multi-player no-limit Texas Hold'em is the only AI that has reached superhuman level in any large game beyond two-player zero-sum games.
It is completely based on game-theoretic principles\footnote{
Pluribus had many approximations, and approximated an even weaker game-theoretic solution concept than Nash equilibrium: coarse correlated equilibrium.
}, and reached superhuman level without any training data.
Many others had been trying to reach superhuman level on that exact problem for 70 years with rule-based approaches, supervised learning, reinforcement learning, \emph{etc.}.
Only the game-theoretic approach succeeded. 
This is despite the game almost surely having a very large or infinite number of equilibria.

Furthermore, finding NE of multiagent systems is useful not only from the perspective of an individual agent deciding which strategy to play, but also from the perspective of social scientists outside the system trying to understand it, as well as social planners trying to direct it.\footnote{
In such situations, it is often the \emph{aggregate behavior} of players that matters, not which specific, individual players decide to play which specific strategies.
}
For example, in the field of mechanism design, social planners aim to design games whose equilibria have desired properties.
Our method opens a new way to tackle such problems when there are many---possibly infinitely many---players.

\section*{Acknowledgments}

This material is based on work supported by the Vannevar Bush Faculty Fellowship ONR N00014-23-1-2876, NSF grants RI-2312342 and RI-1901403, ARO award W911NF2210266, and NIH award A240108S001.

\bibliographystyle{plainnat}
\bibliography{dairefs,references}
\end{document}